# A Probabilistic Approach for Maintaining Trust Based on Evidence

**Yonghong Wang**                                          YHWANG@ANDREW.CMU.EDU
*Robotics Institute*
*Carnegie Mellon University*
*5000 Forbes Ave*
*Pittsburgh, PA 15213 USA*

**Chung-Wei Hang**                                              CHANG@NCSU.EDU
**Munindar P. Singh**                                          SINGH@NCSU.EDU
*Department of Computer Science*
*North Carolina State University*
*Raleigh, NC 27695-8206 USA*

## Abstract

Leading agent-based trust models address two important needs. First, they show how an agent may estimate the trustworthiness of another agent based on prior interactions. Second, they show how agents may share their knowledge in order to cooperatively assess the trustworthiness of others. However, in real-life settings, information relevant to trust is usually obtained piecemeal, not all at once. Unfortunately, the problem of *maintaining* trust has drawn little attention. Existing approaches handle trust updates in a heuristic, not a principled, manner.

This paper builds on a formal model that considers probability and certainty as two dimensions of trust. It proposes a mechanism using which an agent can update the amount of trust it places in other agents on an ongoing basis. This paper shows via simulation that the proposed approach (a) provides accurate estimates of the trustworthiness of agents that change behavior frequently; and (b) captures the dynamic behavior of the agents. This paper includes an evaluation based on a real dataset drawn from Amazon Marketplace, a leading e-commerce site.

## 1. Introduction

Let us consider applications in domains such as electronic commerce, social networks, collaborative games, and virtual worlds populated with multiple virtual characters. These applications exhibit two important common features: (1) they naturally involve multiple entities, real (humans or businesses) or fictional; and (2) these entities are—or behave as if they are—autonomous and heterogeneous. For this reason, we view these entities or their computational surrogates as *agents*. The success of an agent application, evaluated in terms such as the quality of experience enjoyed by a user or the economic value derived by a business, depends on felicitous interactions among the agents. Since the agents are functionally autonomous, the felicity of their interactions cannot be centrally ensured. Further, each agent usually has limited knowledge of the others with whom it interacts. Therefore, each agent relies upon a notion of *trust* to identify agents with whom to interact.

Given our intended applications, we narrow our scope to agents who not only provide and consume services, but also share information regarding the trustworthiness of other agents. We assume each agent behaves according to a fixed type, meaning that although its behavior could be complex, its trustworthiness is not based on the incentives or sanctions it might receive, and its behavior is not different toward different participants. One can imagine settings such as service encounters where a service provider does not selectively favor some of its customers. Hence, the purpose of the trust model is to distinguish good from bad agents, not directly to cause agents to behave in a good manner. Further, we assume that the setting is *empirical*, meaning that the agents base the extent of their trust in others upon the outcomes of prior interactions. The level of trust an agent Alice places in an agent Bob can be viewed as Alice's prediction of Bob providing it a good service outcome in the future. To be empirically reliable, Alice should estimate Bob's trustworthiness based on its past experience with Bob. Both because (1) the parties with whom an agent deals may alter their behavior and (2) the agent receives information about other parties incrementally, it is important that an agent be able to update its assessments of trust.

As one would expect with such an important subject, several researchers have developed formal ways to represent and reason with trust. Interestingly, however, existing approaches do not concentrate on how to *maintain* such representations. It might seem that researchers believe that a heuristic approach would be adequate. The typical approach is based on exponential discounting, and requires a programmer to hand-tune parameters such as a discount factor.

This paper contributes a model and method for updating trust ratings in light of incremental evidence. Specifically, it develops a principled, mathematical approach for maintaining trust *historically* (as a way to evaluate agents who provide services) and *socially* (as a way to evaluate agents who provide information about other agents). Further, this paper shows how to avoid any hand-tuned parameter.

## 1.1 Technical Motivation

A common way to estimate the trustworthiness of a provider is to evaluate the probability of a future service outcome being good based on the number of good service outcomes from the provider in the past. However, a traditional scalar representation (i.e., a probability) cannot distinguish between getting one good outcome from two interactions, and getting 100 good outcomes from 200 interactions. But, intuitively, there is a significant difference in terms of the confidence one would place in each of the above two scenarios. For this reason, modern trust models define trust in terms of both the probability and the certainty of a good outcome (Jøsang & Ismail, 2002; Wang & Singh, 2007; Gómez, Carbó, & Earle, 2007; Teacy, Patel, Jennings, & Luck, 2006; Harbers, Verbrugge, Sierra, & Debenham, 2007; Paradesi, Doshi, & Swaika, 2009). The certainty is a measure of the confidence that an agent may place in the trust information. Computing the certainty can help an agent filter out those parties for whom it has insufficient information, even if nominally the probability of a good outcome is high. In general, the certainty of a trust value should (a) increase as the amount of information increases with a fixed probability, and (b) decrease as the number of conflicts increases with a fixed total number of experiences (Wang & Singh, 2010).

Open systems are dynamic and distributed. In other words, an agent often needs to select a service provider with whom it has had no previous interaction. *Referral networks* enable agents to collect trust information about service providers in a distributed manner (Yu & Singh, 2002; Procaccia, Bachrach, & Rosenschein, 2007). In a referral network, an agent requests other agents, called *referrers*, to provide trust information about a service provider. If a referrer lacks direct experience with the service provider, it may refer to another (prospective) referrer. Existing trust models (e.g., Barber & Kim, 2001), specify how an agent may aggregate trust information from multiple sources (which could include a combination of referrals and direct interactions).

We view referrals themselves as services that the referrers provide. Consequently, an agent ought to be able to estimate a referrer's trustworthiness based on the quality of the referrals it provides. However, existing trust models lack a principled mechanism by which to *update* the trust placed in a referrer.

Besides, to reflect the dynamism of agents over time, a *discount factor* is needed to help trust models provide accurate predictions of future behavior (Zacharia & Maes, 2000; Huynh, Jennings, & Shadbolt, 2006). With a low discount factor, past behavior is forgotten quickly and the estimated trustworthiness reflects recent behavior. Conversely, with a high discount factor, the estimated trustworthiness considers and emphasizes the long-term overall behavior of the rated agent. Different discount factors can yield different accuracy of behavior predictions. Choosing a proper discount factor for different types of agents in varied settings involves a crucial trade-off between accuracy and evidence. This trade-off, however, has not drawn much attention in the trust research community.

We propose a probabilistic approach for updating trust that builds on Wang and Singh's (2010) probability-certainty trust model. Our trust update method enriches Wang and Singh's trust model in two ways. First, our trust update applies in estimating the trustworthiness of referrers based on the referrals they provide. Second, our method adjusts the discount factor *dynamically* by updating the dynamism of the agents without requiring any manual tuning.

We select Wang and Singh's trust model because it supports some features that are crucial for our purposes. One, it defines the trustworthiness of an agent in evidence space, representing trust using both probability and certainty. Two, it defines certainty so that the amount of trust placed in an agent increases with the amount of evidence (if the extent of conflict is held constant) and decreases with increasing conflict (if the amount of evidence is held constant). Three, it supports operators for propagating trust through referrals. Wang and Singh (2006) define mathematical operators for propagating trust. We incorporate these operators as bases for addressing the specific technical problems of computing trust updates and discounting referrers who provide erroneous referrals.

## 1.2 Contributions

This paper proposes a principled, evidence-based approach by which an agent can update the amount of trust it places in another agent. It introduces formal definitions for updating the trust placed and studies their mathematical properties. To achieve a self-tuning approach for trust updates, this paper proposes the new notion of *trust in history*, in contrast to the traditional notion of discounting history via a hand-tuned discount factor. This paper evaluates the proposed approach (1) conceptually via comparison with existing approaches in terms of formal properties (2) via simulation against different agent behavior profiles; and (3) with respect to some data from a real-life marketplace. The main outcomes are that our approach

- Does not require the fine-tuning of parameters by hand, thereby not only reducing the burden on a system administrator or programmer, but also expanding the range of potential applications to include those where the behavior profiles of the agents are not known ahead of time.

- Yields precision in estimating the probability component of trust.

- Yields a more appropriate level of the certainty component of trust than existing approaches. In particular, it recognizes the effect of conflict in evidence and how to compute certainty based on the certainty of the input information.

- Is robust against agents who provide wrong information.

### 1.3 Organization

The rest of this paper is organized as follows. Section 2 provides the essential technical background for our approach. Section 3 introduces a general model for trust update that uniformly handles both historical and social updates. Section 4 introduces a series of trust update methods culminating in our proposed method. Section 5 evaluates these methods on theoretical grounds by establishing theorems regarding the desirable and undesirable properties. Section 6 specifies the historical and social update scenarios precisely. Section 7 conducts an extensive experimental evaluation of our methods, including both simulations and an evaluation using real marketplace data from Amazon. Section 8 studies the literature. Section 9 concludes with a discussion and some directions for future work. Appendix A presents proofs for all theorems.

## 2. Background on Probabilistic Trust Representation

This section introduces the key background on Wang and Singh's (2007) approach that is necessary for understanding our present contribution.

### 2.1 Probability-Certainty Distribution Function

Considering a binary event $\langle r, s \rangle$, where $r$ and $s$ represent the number of positive and negative outcomes, respectively. Let $x \in [0, 1]$ be the probability of a positive outcome. Then the posterior probability of evidence $\langle r, s \rangle$ is the conditional probability of $x$ given $\langle r, s \rangle$ (Casella & Berger, 1990). The conditional probability of $x$ given $\langle r, s \rangle$ is

$$f(x|\langle r, s \rangle) \quad = \frac{g(\langle r, s \rangle|x)f(x)}{\int_0^1 g(\langle r, s \rangle|x)f(x)dx}$$
$$= \frac{x^r(1-x)^s}{\int_0^1 x^r(1-x)^s dx}$$

where $g(\langle r, s \rangle | x) = \begin{pmatrix} r + s \\ r \end{pmatrix} x^r (1 - x)^s$.

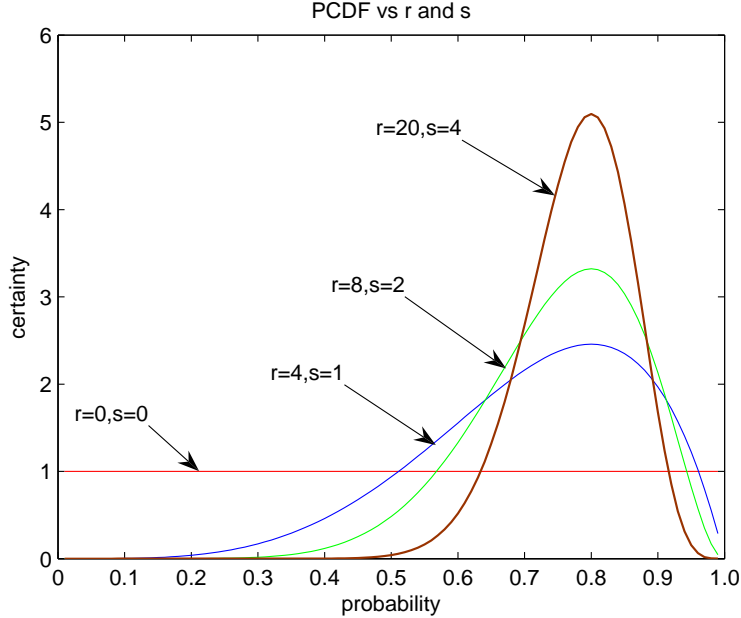

Figure 1: Examples of probability-certainty distribution functions, varying $r$ and $s$.

Here $f(x)$ is the probability distribution function of $x$, which is itself the probability of a positive outcome. The signature of $f$ is given by $f : [0, 1] \mapsto [0, \infty)$. Because $f$ is a probability density, we have $\int_0^1 f(x)dx = 1$. Following Jøsang (2001), we interpret the above as a probability of a probability or a *probability-certainty distribution function (PCDF)*. The probability that the probability of a positive outcome lies in $[x_1, x_2]$ equals $\int_{x_1}^{x_2} f(x)dx$. The mean value of $f$ is $\frac{\int_0^1 f(x)dx}{1-0} = 1$. Figure 1 gives some examples of $f(x)$ for different numbers of positive and negative outcomes. Notice that when we have no evidence (i.e., $\langle r, s \rangle = \langle 0, 0 \rangle$), we obtain a uniform distribution. As the evidence mounts, the distribution becomes more and more focused around its expected value.

As an aside, notice that although we consider integral values of $r$ and $s$ in the above examples, the actual values of $r$ and $s$ would usually not be integral because of the effect of discounting information received from others or remembered from past interactions. In particular, it is possible that the total evidence is positive but less than one, i.e., $0 < r + s < 1$.

## 2.2 Trust Representation

As in Jøsang's approach, Wang and Singh's model represents trust values in both the evidence and the belief spaces. In evidence space, a trust value is in the form $\langle r, s \rangle$, where $r + s > 0$. Here, $r \geq 0$ is the number of positive experiences (that, say, agent Alice has with agent Bob) and $s \geq 0$ is the number of negative experiences she has with Bob. Both $r$ and $s$ are real numbers. Given $\langle r, s \rangle$, $\alpha = \frac{r}{r+s}$ is the expected value of the probability of a positive outcome when $r + s > 0$ and can be set as $\alpha = 0.5$ when $r + s = 0$. In belief space, a trust value is modeled as a triple of belief, disbelief, and uncertainty weights, $\langle b, d, u \rangle$, where each of $b$, $d$, and $u$ is greater than 0 and $b + d + u = 1$. In intuitive terms, the certainty $c = 1 - u$ represents the confidence placed in the probability. Trust values can be translated between the evidence and belief space.

Wang and Singh (2007) differ from Jøsang (2001) in their definition of certainty. Wang and Singh's definition is based on the following intuition. As Figure 1 shows for $\langle r, s \rangle = \langle 0, 0 \rangle$ when we know nothing, $f$ is a uniform distribution over probabilities $x$. That is, $f(x) = 1$ for $x \in [0, 1]$ and 0 elsewhere. This reflects the Bayesian intuition of assuming an equiprobable prior. Intuitively, the uniform distribution has a certainty of 0. As additional knowledge is acquired, the probability mass shifts so that $f(x)$ is above 1 for some values of $x$ and below 1 for other values of $x$. For the above reason, Wang and Singh (2007) define certainty to be the area above the uniform distribution $f(x) = 1$.

**Definition 1** *The certainty based on evidence $\langle r, s \rangle$, is given by*

$$
\begin{aligned}
\boldsymbol{c}(r, s) &= \frac{1}{2} \int_0^1 |f(x) - 1| dx \\
&= \frac{1}{2} \int_0^1 \left| \frac{x^r(1-x)^s}{\int_0^1 x^r(1-x)^s dx} - 1 \right| dx
\end{aligned}
$$

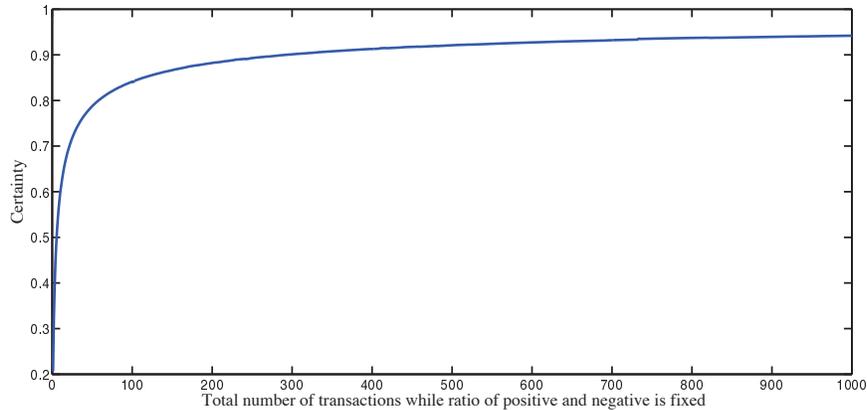

Figure 2: Certainty increases with mounting evidence provided the amount of conflict in the evidence is held constant. The X-axis measures the total number of outcomes, which are equally positive and negative.

Conflict in the evidence in this setting means that some evidence is positive and some is negative. Thus conflict is maximized when $r = s$ and is minimized when $r$ or $s$ is zero. Wang and Singh (2007) prove that certainty increases when the total number of transactions increases and the conflict is fixed, as in Figure 2. They also show that certainty decreases when conflict increases and the total number of transactions is fixed, as in Figure 3.

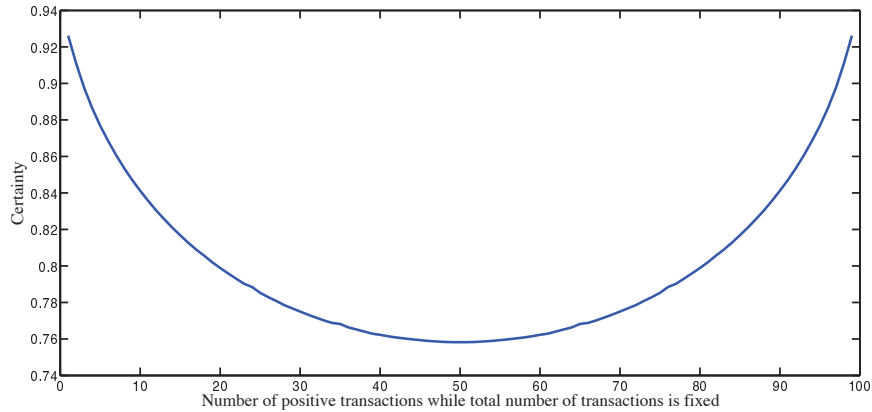

Figure 3: Certainty decreases with increasing conflict provided the amount of evidence is held constant. The X-axis measures the number of positive outcomes out of a fixed total number of outcomes.

## 2.3 Trust Propagation

In real-life settings, an agent (a prospective client) may lack direct experience with another agent (a prospective service provider) with whom it considers interacting. In this case, the client can ask referrers for trust referrals. If a referrer lacks direct experience, it may refer to other referrers, and so on. This is the essential idea behind referral networks. But how should we calculate trust through referral networks? Many researchers have studied trust propagation. In our chosen framework, Wang and Singh (2006) define mathematical operators for propagating trust, which we can leverage for our present goals.

Wang and Singh (2006) provide a concatenation operator (similar to Jøsang's, 1998, recommendation operator) that enables a client $C$ to compute how much trust it should place in a service provider $S$ based on its direct experience with a referrer $R$ and a referral for $S$ provided by $R$. The idea is that, to compute its trust in $S$, $C$ simply concatenates its trust in $R$ with $R$'s report about $S$. Definition 2 captures Wang and Singh's concatenation operator. In our setting, let $M_R = \langle r_R, s_R \rangle$ be agent $C$'s trust in a referrer $R$. Here, $c_R$ is the certainty determined from the above trust value. Further, let $M_S = \langle r', s' \rangle$ be $R$'s report about its trust in a provider $S$. Then the amount of trust to be placed by $C$ in $S$ is given by $M_R \otimes M_S$.

**Definition 2** *Concatenation $\otimes$. Let $M_R = \langle b_R, d_R, u_R \rangle$ and $M_S = \langle b', d', u' \rangle$ be two trust values. Then $M_R \otimes M_S = \langle b_R b', b_R d', 1 - b_R b' - b_R d' \rangle$.*

To handle the situation where $C$ collects trust information of $S$ from more than one source, we use Jøsang's aggregation operator (Jøsang, 2001; Wang & Singh, 2007), which simply sums the available evidence pro and con. $C$ can use this operator to combine independent reports about the trust to place in $S$. Definition 3 captures the aggregation operator. In our setting, $M_i = \langle r_i, s_i \rangle$ would be the trust that $C$ would place in $S$ based on exactly one path from $C$ to $S$. When the paths are mutually independent, i.e., nonoverlapping, $C$'s aggregate trust in $S$ would be given $M_1 \oplus \ldots \oplus M_k$.

**Definition 3** *Aggregation* $\oplus$. *Let* $M_1 = \langle r_1, s_1 \rangle$ *and* $M_2 = \langle r_2, s_2 \rangle$ *be two trust values. Then,* $M_1 \oplus M_2 = \langle r_1 + r_2, s_1 + s_2 \rangle$.

## 3. General Model for Updating Trust

As we observed above, most existing trust models do not provide a suitable trust update method by which an agent may maintain the trust it places in another agent. In our model, trust updates arise in two major settings, which we consolidate into a universal model for trust update. These settings are as follows.

- *Trust update for referrers*, wherein an agent updates the trust it places in a referrer based on how accurate its referrals are. This is a way for an agent to maintain its social relationship with a referrer.

- *Trust update by trust in history*, wherein an agent updates the trust it places in a service provider by tuning the relative weight (discount factor) assigned to the service provider's past behavior with respect to its current behavior. This is a way for an agent to accommodate the dynamism of a service provider.

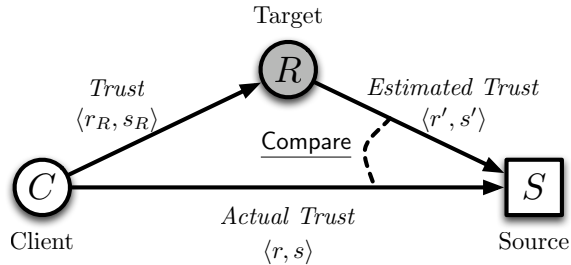

Figure 4: Schematic illustration of our generalized trust update approach. Throughout this paper, we use $\alpha = \frac{r}{r+s}$, $\alpha' = \frac{r'}{r'+s'}$, and $\alpha_R = \frac{r_R}{r_R+s_R}$

Interestingly, our approach treats the above settings as variations on a common theme, which we term our general model for updating trust. Figure 4 presents this model, which summarizes a process consisting of the following steps based on a client $C$, a target $R$, and a service provider $S$. The client $C$ seeks to update the amount of trust $\langle r_R, s_R \rangle$ that $C$ places in a target $R$.

- $C$ estimates $R$'s trustworthiness (before the update) as $\langle r_R, s_R \rangle$.

- $R$ reports $S$'s trustworthiness as $\langle r', s' \rangle$.

- $S$ delivers an outcome from which $C$ obtains direct information by which it can estimate the actual trustworthiness $\langle r, s \rangle$ of $S$.

- Using this information about the (apparent) trustworthiness of $S$, $C$ determines the accuracy of the estimated trust value it previously received from $R$. Based on this measure of $R$'s accuracy, $C$ updates the trust it places in $R$ to $\langle r'_R, s'_R \rangle$ on empirical grounds.

Trust update methods can be differentiated by how they compare the estimated and actual trust values. The rest of this section discusses the general structure of trust update and investigates some trust update methods along with their shortcomings. Section 4 introduces our preferred approach.

Let us follow the setting of Figure 4. The accuracy of the estimated trust value is defined by the closeness of the estimated trust value $\langle r', s' \rangle$ and the actual trust value $\langle r, s \rangle$. Instead of interpreting each transaction (one estimation of trust from the target) as either good or bad, we interpret it as $q$ good and $1 - q$ bad transactions. Notice that "good" and "bad" in the present context of a target providing estimates refers to the accuracy or otherwise of the estimates with respect to the actual outcomes the client receives from the service provider. Thus a target's estimate could be good or bad independently of whether the actual service outcome is good or bad.

Thus, $q$ reflects how close the estimation $\langle r', s' \rangle$ is to the actual trust $\langle r, s \rangle$. We require $0 \leq q \leq 1$ ranging from a perfectly inaccurate to a perfectly accurate referral. The weight we assign such an estimation should increase with its certainty. For example, suppose the actual trustworthiness of $S$ is $\langle 10, 0 \rangle$. Say, one target $R_A$ estimates $S$'s trustworthiness as $\langle 0, 1 \rangle$ and another target $R_B$ estimates it as $\langle 0, 100 \rangle$. Both estimates agree that $S$ is not trustworthy, but $R_A$ claims much lower certainty ($\mathbf{c}(0, 1) = 0.25$) than does $R_B$ ($\mathbf{c}(0, 100) = 0.99$). So $R_A$ should be punished less than $R_B$ if both estimates turn out to be inaccurate. And, similarly, for rewarding them in case of accuracy. Therefore, instead of treating each transaction as one transaction, we treat it as $c'$ transactions, where $c' < 1$ is the certainty in the estimation. That is, we interpret each estimation as $c'q$ good and $c'(1 - q)$ bad transactions.

In addition, we discount each past transaction by its age (Zacharia & Maes, 2000; ?, ?, ?, ?, ?). Now let $\langle r_R, s_R \rangle$ be the trust placed in $R$ by $C$; $\langle r', s' \rangle$ be an estimation with $c' = \mathbf{c}(r', s')$; and $\langle r'_R, s'_R \rangle$ be the updated trust placed in $R$ by $C$. Algorithm 1 presents a *modular* specification for our generic trust update approach, highlighting its key inputs and outputs. Let $\beta$ be the temporal discount factor. Based on actual observations $\langle r, s \rangle$, let $q$ represent how accurate the estimation is and $p$ represent how bad the estimation is. The specific approaches that we consider below differ in how each computes its measure of accuracy, $q$.

Note that we assume the client updates its trust in the referral after it has conducted some transactions with the service provider, so $r + s > 0$ in this case. When $\langle r, s \rangle = \langle 0, 0 \rangle$, the client cannot update the trust according to our formulation, since we discount the update by the certainty of $\langle r, s \rangle$ and when $\langle r, s \rangle = \langle 0, 0 \rangle$, the certainty is 0. It does not matter if $\langle r_R, s_R \rangle = 0$ in the update method. We can initialize $\langle r_R, s_R \rangle$ to be some predefined number, for example, $\langle 1, 1 \rangle$ or $\langle 0, 0 \rangle$. The initial value corresponds to the prior distribution of the trust. The uniform distribution corresponds to initial setting of $\langle r_R, s_R \rangle$ at $\langle 0, 0 \rangle$. But we can also set any prior we deem fit for the system. In trust update for referrers, we initialize $\langle r_R, s_R \rangle$ to $\langle 1, 1 \rangle$ (to suggest that the client is willing to consider a referral from

---

**Algorithm 1**: generalUpdate: Abstract method to revise the trust placed in target $R$.

---

**input** $q$, $p$, $\beta$, $c'$, $\langle r_R, s_R \rangle$;
$\delta r_R \leftarrow c'q$;
$\delta s_R \leftarrow c'p$;
$r'_R \leftarrow \delta r_R + (1-\beta)r_R$;
$s'_R \leftarrow \delta s_R + (1-\beta)s_R$;
**return** $\langle r'_R, s'_R \rangle$;

---

the referrer) and $\langle r, s \rangle$ to $\langle 0, 0 \rangle$ (to suggest that the client has no prior experience with the provider). But we perform a trust update only after the client has done some transactions with the service provider (i.e., $r + s > 0$). In the trust in history setting, we initialize $\langle r_R, s_R \rangle$ to $\langle 0.9, 0.1 \rangle$ (to suggest that the client trust its own past experience with little confidence) and $\langle r, s \rangle$ to $\langle 0, 0 \rangle$ (to suggest that the client has no prior experience with the provider).

Now let us consider the certainty density function based on $\langle r, s \rangle$, which reflects the actual trustworthiness of $S$. Here, $f(x)$ is the probability (density) of the quality of service provided by $S$ being $x$.

$$f(x) = \frac{x^r(1-x)^s}{\int_0^1 x^r(1-x)^s dx} \tag{1}$$

This density maximizes at $x = \alpha$ meaning that $S$ will most likely provide a service outcome with a quality of $\alpha$. Consider the probability density function based on the estimation $\langle r', s' \rangle$, which is $R$'s estimate of $S$'s trustworthiness and reflects $R$'s assessment of the quality $S$ will provide. In other words, $R$ expects that $S$ will provide a service outcome most likely with a quality of $x = \alpha'$.

## 4. Trust Update Based on Average Accuracy

Based on the above background, we now study the trust update problem systematically. We analyze the shortcomings of a series of approaches, culminating in an approach that yields the characteristics we desire.

### 4.1 Linear Update and Its Shortcomings

*Linear* is a common trust update method, which serves as a baseline for comparison. *Linear* defines the accuracy as the absolute difference between the quality $\alpha'$ of the estimated trust value $\langle r', s' \rangle$ and the quality $\alpha$ of the actual trust value $\langle r, s \rangle$. That is,

$$q = 1 - |\alpha - \alpha'| \tag{2}$$

Using Equation 2, we construct two trust update methods, *Linear-WS* and *Jøsang*, by inserting the $q$ defined in *Linear* into the general trust update model described in Algorithm 1. Note that, *Linear-WS* and *Jøsang* use trust representations with their separate definitions of certainty, thereby yielding different trust update methods. *Linear-WS*, as shown in Algorithm 2, adopts Wang and Singh's notion of certainty underlying trust (Definition 1), whereas Algorithm 3 depicts *Jøsang* trust update method, where the certainty $c'$ is defined as $\frac{r'+s'}{r'+s'+2}$ (Jøsang, 2001). Jøsang (1998) defines certainty as $\frac{r'+s'}{r'+s'+1}$, which yields little difference from a later work by Jøsang (2001) in trust update. Our motivation for introducing *Linear-WS* is to help distinguish the benefit of our trust update methods from the benefit of using Wang and Singh's static trust representation. As we show below, *Linear-WS* (which combines Wang and Singh's static model with a heuristic update) performs worse than our proposed update methods, thereby establishing that the proposed methods yield some benefits beyond the static model that they incorporate.

A shortcoming of *Linear* is that it does not consider certainty. Consider an agent who reports $\langle 0.1, 0.1 \rangle$ about a service provider when it has little information about the provider and reports $\langle 90, 10 \rangle$ later when it has gathered additional information. Suppose the service provider's quality of service is indeed 0.9. The the above agent should be accorded high trust, since it reports correct trust value of the service provider with high certainty and reports wrong trust value with low certainty. In other words, when updating the trust in this agent, the second referral should be given more weight than the first one. However, *Linear* treats both referrals as the same and thus ends up with a wrong updated trust value of this agent.

---

**Algorithm 2**: Linear-WS: A trust update method to revise the trust placed in target $R$.

**input** $\langle r, s \rangle$, $\langle r', s' \rangle$, $\beta$, $\langle r_R, s_R \rangle$;

$$\alpha \leftarrow \frac{r}{r+s}$$

$$\alpha' \leftarrow \frac{r'}{r'+s'}$$

$$q \leftarrow 1 - |\alpha - \alpha'|$$

$$c' \leftarrow \mathbf{c}(r', s')$$

**return** *generalUpdate*$(q, 1-q, \beta, c', \langle r_R, s_R \rangle)$;

---

**Algorithm 3**: Jøsang: A trust update method to revise trust placed in agent $R$.

**input** $\langle r, s \rangle$, $\langle r', s' \rangle$, $\beta$, $\langle r_R, s_R \rangle$;

$$\alpha \leftarrow \frac{r+1}{r+s+2}$$

$$\alpha' \leftarrow \frac{r'+1}{r'+s'+2}$$

$$q \leftarrow 1 - |\alpha - \alpha'|$$

$$c' \leftarrow \frac{r'+s'}{r'+s'+2}$$

**return** *generalUpdate*$(q, 1-q, \beta, c', \langle r_R, s_R \rangle)$;

## 4.2 Update Based on Max-Certainty and Its Shortcomings

Hang, Wang, and Singh (2008) proposed the *Max-Certainty* trust update method, which applies on Wang and Singh's trust representation. *Max-Certainty* supports some interesting features, but suffers from some shortcomings, which the present approach avoids. *Max-Certainty* follows the general update model of Section 3, and defines $q$ as in Algorithm 4.

---

**Algorithm 4**: Max-Certainty: A trust update method to revise the trust placed in target $R$.

**input** $\langle r, s \rangle$, $\langle r', s' \rangle$, $\beta$, $\langle r_R, s_R \rangle$;

$$\alpha \leftarrow \frac{r}{r + s}$$

$$\alpha' \leftarrow \frac{r'}{r' + s'}$$

$$q \leftarrow \frac{\alpha'^r (1 - \alpha')^s}{\alpha^r (1 - \alpha)^s}$$

$$c' \leftarrow \mathbf{c}(r', s')$$

**return** *generalUpdate*$(q, 1 - q, \beta, c', \langle r_R, s_R \rangle)$;

---

The intuition behind Algorithm 4's definition of $q$ is as follows. From Equation 1, we have that $q = \frac{f(\alpha')}{f(\alpha)}$, since $S$ will provide service most likely with a quality of $\alpha$—and not as likely at any other quality $\alpha'$. That is, $q$ measures the ratio of the likelihoods that the service provided by $S$ has qualities $\alpha'$ and $\alpha$, respectively. In other words, our measure of accuracy $q$ is the ratio of the probability computed from the estimate from $R$ with respect to the probability computed from the measurement made by $C$ itself. Figure 5 illustrates this computation.

A *malicious* target (such as a referrer) in the present setting is one who is not just wrong but is over-confident—that is, such an agent exaggerates the amount of evidence it claims behind its report about a service provider. Although *Max-Certainty* tells us how likely it is that a target $R$ is trustworthy, it is not sensitive against malicious targets. Since combined trust is naturally weighed by the amount of evidence, a trust report from a malicious target may falsely dominate truthful reports based on an apparently smaller amount of evidence.

Under *Max-Certainty*, it takes $C$ a long time to pinpoint a malicious target. For example, suppose the actual trustworthiness of $S$ is $\langle 2, 1 \rangle$ and $R$ reports $\langle 5, 5 \rangle$. According to *Max-Certainty*, $\langle \delta r_R, \delta s_R \rangle = \langle 0.37, 0.07 \rangle$. If $R$ had reported trust in $S$ of $\langle 1000, 1000 \rangle$, then $\langle \delta r_R, \delta S_R \rangle$ would equal $\langle 0.79, 0.15 \rangle$. In other words, *Max-Certainty* treats the new evidence as being confirmatory. Instead, we claim the report of $\langle 1000, 1000 \rangle$ is bad because it exaggerates the available evidence and thus has a highly misleading effect. In particular, it may end up overriding many accurate reports (of lower certainty). Therefore, we consider the report of $\langle 1000, 1000 \rangle$ to inaccurate, and treat its evidence as being disconfirmatory. This observation leads to the following update approaches.

In the update formula based on *Max-Certainty*, let $R$'s trust of $S$ be $\langle r, s \rangle$. When $r + s$ is large, the formula is highly sensitive, since the PCDF is maximized at $x = \alpha = \frac{r}{r+s}$, but decreases quickly when $x$ deviates from $\alpha$ (in other words, close to $x = \alpha$, the magnitude of the slope is high). For example, let the (actual) trustworthiness of $S$ be $\langle 800, 200 \rangle$. When a referrer reports $\langle 19, 6 \rangle$, it predicts that the quality of $S$ is 0.76. This is quite close to the actual quality of $S$, namely, 0.80. However, *Max-Certainty* yields $\langle \delta r_R, \delta s_R \rangle = \langle 0.06, 0.58 \rangle$, which indicates that *Max-Certainty* treats it to be a poor estimation.

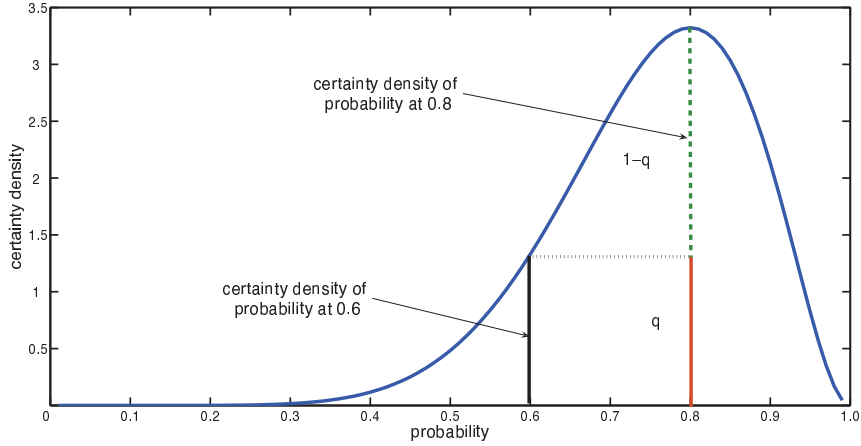

Figure 5: Illustration of the trust update method *Max-Certainty* with $\langle r, s \rangle = \langle 8, 2 \rangle$ and $\langle r', s' \rangle = \langle 6, 4 \rangle$.

### 4.3 Update Based on Sensitivity and Its Shortcomings

The foregoing leads us to another update method, which we term *Sensitivity*. The intuition underlying *Sensitivity* was identified by Teacy et al. (2006). To motivate *Sensitivity*, consider that in $R$'s estimate, the probability that the quality of $S$ equals $p$, is given by

$$l(p) = \frac{p^{r'}(1-p)^{s'}}{\int_0^1 x^{r'}(1-x)^{s'}dx}$$

Clearly, $l(p)$ maximizes at $\alpha'$, which means that $R$ estimates the quality of the service provided by $S$ as most likely being $\alpha'$. If we normalize $l(\alpha')$ to 1, then

$$q = \frac{\alpha^{r'}(1-\alpha)^{s'}}{\alpha'^{r'}(1-\alpha')^{s'}} \tag{3}$$

measures how likely in $R$'s assessment would the quality of the service provider be $\alpha$.

Figure 6 illustrates the above formulas using $\langle r, s \rangle = \langle 8, 2 \rangle$ and $\langle r', s' \rangle = \langle 6, 4 \rangle$. Returning to the previous example, $\langle \delta r_R, \delta s_R \rangle = \langle 0.01, 0.93 \rangle$, which means that *Sensitivity* considers $\langle 1000, 1000 \rangle$ an inaccurate report—as it should.

Although *Sensitivity* improves over *Max-Certainty*, like *Max-Certainty*, it remains susceptible to being excessively sensitive when the number of transactions is large. A numerical example of the susceptibility of *Sensitivity* is as presented in Section 4.2. Let the trustworthiness of $S$ be $\langle 800, 200 \rangle$, whereas the referrer reports $\langle 190, 60 \rangle$. The above method, which incorporates uncertainty, yields $\langle \delta r_R, \delta s_R \rangle = \langle 0.24, 0.55 \rangle$. In other words, it treats the referral as being bad. However, $190/(190 + 60) = 0.76$ is quite close to $800/(800 + 200) = 0.80$, which indicates that it ought to be treated as a good referral—hence we have a discrepancy. The problem with *Sensitivity* is that it treats a referral as being disconfirmed by evidence simply because the referrer is too confident even though the referral is accurate. Theorem 3 of Section 5 demonstrates the above problem, in a general setting.

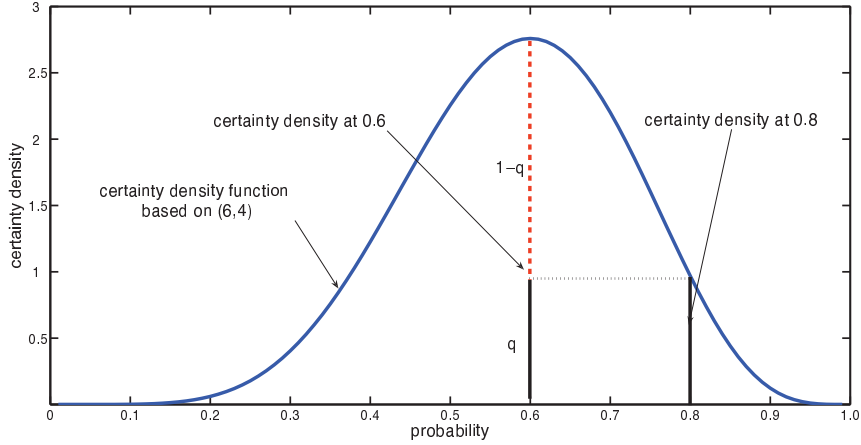

Figure 6: Illustration of the trust update method based on *Sensitivity* with $\langle r, s \rangle = \langle 8, 2 \rangle$ and $\langle r', s' \rangle = \langle 6, 4 \rangle$. Thus $\alpha = 0.8$ and $\alpha' = 0.6$.

The bottom-line is that both *Max-Certainty* and *Sensitivity* produce undesirable results.

## 4.4 Average Accuracy Disregarding Uncertainty

The foregoing discussion leads us to our penultimate step in coming up with our desired approach. This and the final approach (detailed in Section 4.5) both consider the average accuracy of the estimations. The present variant is the simpler of the two because it disregards the uncertainty inherent in the belief regarding the source $S$.

Suppose as an idealization that the actual trustworthiness of the source $S$ equals $\langle \alpha, 1 - \alpha, 0 \rangle$, when expressed as a belief-disbelief-uncertainty triple. This case arises when we know for sure that the source can provide the quality of service or referral at probability $\alpha$. Therefore, the uncertainty is 0, indicating that this is an ideal case.

As Figure 7 shows, suppose the target reports a trust value about the service provider of $\langle r', s' \rangle$. Let $c' = \mathbf{c}(r', s')$ be the certainty based on $\langle r', s' \rangle$. The PCDF for $\langle r', s' \rangle$ means the target estimates that the provider will produce a quality of $x \in (0, 1)$ with certainty $f(x)$:

$$f(x) = \frac{x^{r'}(1-x)^{s'}}{\int_0^1 x^{r'}(1-x)^{s'}dx}$$

But the provider's actual quality is $x = \alpha$. The square of the estimation error at $x$ is $(x - \alpha)^2$. We multiply the square of the error by its certainty $f(x)$, and then integrate it over $x$ from 0 to 1 to obtain the average square of the estimation error. The square root of this integration yields the average error. That is, we can calculate the error in the estimation according to the following formula:

$$e = \sqrt{\frac{\int_0^1 x^{r'}(1-x)^{s'}(x-\alpha)^2 dx}{\int_0^1 x^{r'}(1-x)^{s'} dx}}$$

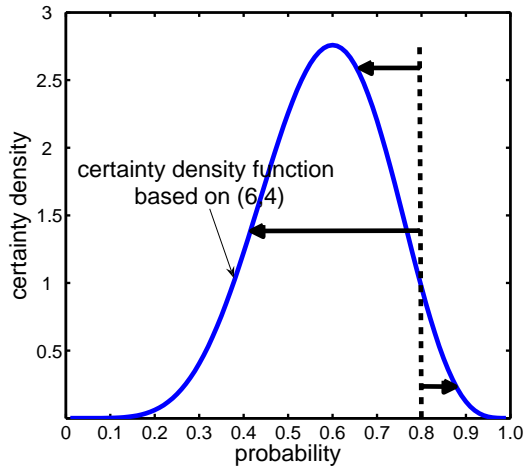

Figure 7: Illustration of the average trust update method for $r' = 6$ and $s' = 4$, and $\alpha = 0.8$ (dashed line). The error $e$ is the average of the length of the arrows.

There are many ways to compute average errors, including the $L_1$ and $L_\infty$ norms, and so on. We use the $L_2$ norm here for its simple mathematical properties. This choice is not unique but is a common one (the same as variance) and is convenient to manipulate mathematically.

We can now give an alternative definition of $q$ based on $e$, i.e., $q = 1 - e$. From the PCDF corresponding to the estimation $\langle r_R, s_R \rangle$, we can see that $q$ corresponds to the average accuracy of the estimation. The updated estimate of trustworthiness $\langle r_R, s_R \rangle$ of $R$ is based on $q$ in the usual manner of Algorithm 1.

## 4.5 Average Accuracy Incorporating Uncertainty

Let us now consider a more complex variant of the above method, which uses the same $q$ definition as the above, but explicitly incorporates the uncertainty inherent in the belief regarding $S$. Treating the trust placed in $S$ as a belief function with uncertainty corresponding to $\langle r, s \rangle$ in evidence space, we would like to discount the updates to $r_R$ and $s_R$ by an additional factor of the certainty of the actual observations $\langle r, s \rangle$ made by $C$. In other

---

**Algorithm 5:** Average-$\beta$: A trust update method to revise trust placed in agent $R$.

**input** $\langle r, s \rangle$, $\langle r', s' \rangle$, $\beta$, $\langle r_R, s_R \rangle$;

$$\alpha \leftarrow \frac{r}{r+s}$$

$$c \leftarrow \mathbf{c}(r, s)$$

$$q = 1 - \sqrt{(\alpha - \frac{r'+1}{r'+s'+2})^2 + \frac{(r'+1)(s'+1)}{(r'+s'+2)^2(r'+s'+3)}}$$

$$c' \leftarrow \mathbf{c}(r', s')$$

$$c' = cc'$$

**return** *generalUpdate*$(q, 1-q, \beta, c', \langle r_R, s_R \rangle)$;

---

words, we would begin with the definition of $q$ as above and discount it with the certainty determined from $\langle r, s \rangle$. Algorithm 5 captures this intuition.

The reason we consider certainty is that when we are not certain of the actual quality of $S$, we are not certain of how to evaluate the target's estimation of $S$ either, so we discount the update by the additional factor $c$. Returning to our previous example, let the trustworthiness of $S$ be $\langle 800, 200 \rangle$, whereas the target reports $\langle 19, 6 \rangle$. The above method, which incorporates uncertainty, yields $\langle \delta r_R, \delta s_R \rangle = \langle 0.53, 0.06 \rangle$, which indicates that the above report is a good estimation—as it is supposed to be since $19/(19+6) = 0.76$, which is quite close to $800/(800 + 200) = 0.8$. Therefore, when $\alpha'$ is close to $\alpha$, no matter how large the total number transactions is, this method considers the target's estimation as being confirmative. The above holds true in general, as Theorem 4 of Section 5 shows.

## 4.6 Understanding the Trade Offs

The following tables illustrate the pros and cons of each update method. We compare the accuracy measurements $q$ used in *Max-Certainty* (Algorithm 4), *Sensitivity* (Equation 3), and *Average-$\beta$* (Algorithm 5).

Table 1 summarizes the various situations of interest, especially those that *Max-Certainty* and *Sensitivity* cannot handle well. As explained in Section 4.3, when $r + s$ is large, *Max-Certainty* is highly sensitive, and thus treats a good report as bad. When $r' + s'$ is large, *Sensitivity* is highly sensitive, and also treats a good report as bad.

Table 2 provides numerical examples corresponding to the situations specified in Table 1. In this table, let the actual quality of the service provider be 0.50 and the quality indicated by a referral be 0.55.

## 4.7 Estimating Certainty

We now evaluate the above methods with respect to their ability to infer and track the certainty of the incoming trust reports. Figure 8 compares the $q$ values produced by different trust update methods. Its $Y$-axis is $q$, as calculated by the different trust update methods introduced above. The estimate's quality $\alpha'$ is fixed at 0.55 as $r' + s'$, the amount of evidence in the estimate, increases. The left and right plots show the resulting $q$ with the actual quality $\langle r, s \rangle = \langle 1, 1 \rangle$, and $\langle r, s \rangle = \langle 200, 200 \rangle$, respectively. *Linear* is always high, independent of the certainty of the report. *Max-Certainty* over-estimates $q$ when $r + s$ is low and under-estimates $q$ when $r + s$ is high, and does not vary with certainty. Only *Average*'s estimate of $q$ reflects the certainty of the reports in both cases.

Table 1: Comparing the effectiveness of trust update methods conceptually.

| Case | | Accuracy | | | |
|---|---|---|---|---|---|
| $r + s$ | $r' + s'$ | *Max-Certainty* | *Sensitivity* | *Linear* | *Average-$\beta$* |
| small | small | good | good | fair | good |
| small | large | good | poor | good | good |
| large | small | poor | good | fair | good |
| large | large | poor | poor | good | good |

Table 2: Trust update methods comparison via numerical examples.

| Case | | Accuracy ($q$) | | | |
|---|---|---|---|---|---|
| $\langle r, s \rangle$ | $\langle r', s' \rangle$ | *Max-Certainty* | *Sensitivity* | *Linear* | *Average-$\beta$* |
| $\langle 1, 1 \rangle$ | $\langle 1.1, 0.9 \rangle$ | 0.99 | 0.99 | 0.95 | 0.78 |
| $\langle 1, 1 \rangle$ | $\langle 220, 180 \rangle$ | 0.99 | 0.13 | 0.95 | 0.95 |
| $\langle 200, 200 \rangle$ | $\langle 1.1, 0.9 \rangle$ | 0.13 | 0.99 | 0.95 | 0.78 |
| $\langle 200, 200 \rangle$ | $\langle 220, 180 \rangle$ | 0.13 | 0.13 | 0.95 | 0.95 |

### 4.8 Methods Summarized

Figure 9 illustrates the trust update methods that we compare, including *Jøsang*, *Linear-WS*, *Max-Certainty*, *Sensitivity*, and *Average-$\beta$*. We categorize these methods with respect to their accuracy measurements and their underlying trust representation. Regarding the accuracy measurement technique, *Jøsang* and *Linear-WS* measure accuracy based on the linear approach. Each of *Max-Certainty*, *Sensitivity*, and *Average-$\beta$* defines its specific accuracy measurement. All the approaches other than *Jøsang* follow Wang and Singh's trust representation.

## 5. Theoretical Evaluation of Accuracy Measurement Techniques

This section evaluates the above trust update methods in theoretical terms by consolidating some important technical results. It may be skipped on a first reading. This section seeks to give technical intuitions about its results.

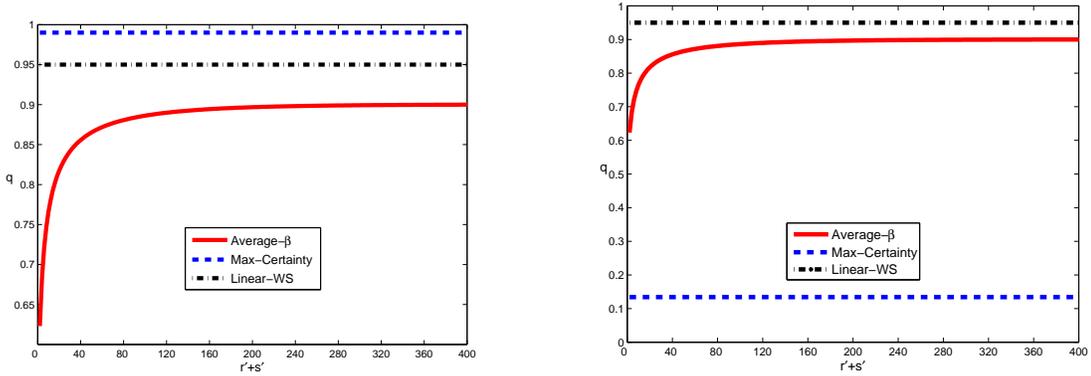

Figure 8: Comparison of trust update methods based on their accuracy measurements $q$. These graphs use a fixed referral expected quality $\alpha' = 0.55$ and vary the amount of reported evidence $r' + s'$. The actual quality values $\langle r, s \rangle$ are low (set to $\langle 1, 1 \rangle$ in the left graph) and high (set to $\langle 200, 200 \rangle$ in the right graph), respectively.

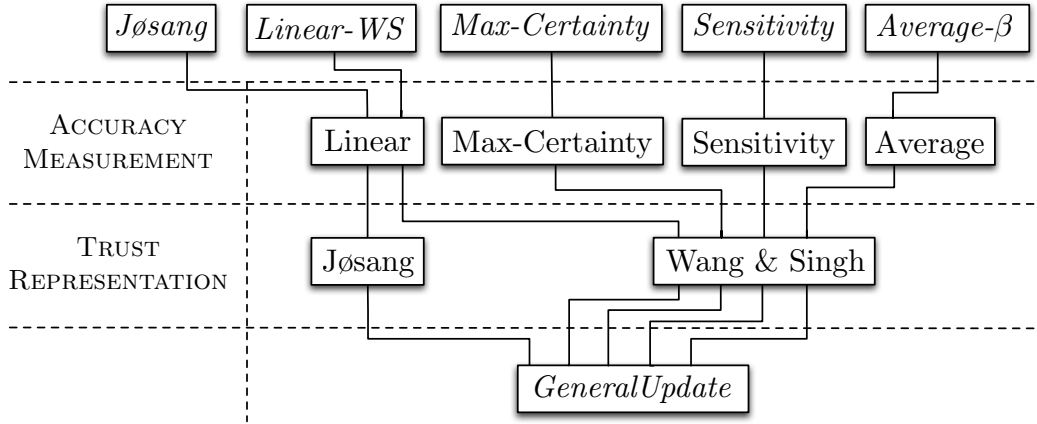

Figure 9: Trust update methods specified in terms of their trust representation and their accuracy measurement methods.

As Figure 9 shows, each trust update method has three main components. The accuracy measurement technique is the main contribution of this paper, and the one we evaluate theoretically.

## 5.1 Bounded Range

As explained above, the update $q$ means that an estimate can be treated as $q$ good and $1 - q$ bad transactions. The range being bounded merely serves as a sanity check on the definitions. Theorem 1 establishes this for all of the accuracy measurement definitions that we consider.

**Definition 4** *Let a trust update method compute q based on the above description. We say the trust update method is* bounded *if and only if, for all inputs,*

$$0 \leq q \leq 1$$

**Theorem 1** *Each of the four definitions of accuracy q as given in Equation 2, Algorithm 4, Algorithm 5, and Equation 3 satisfies boundedness.*

**Proof**: The range is trivially bounded for *Linear*. For *Max-Certainty* and *Sensitivity* methods, we show that the PCDF function, $f(\cdot)$ achieves its maximum at $x = \alpha = \frac{r}{r+s}$, so the $\frac{f(x)}{f(\alpha)}$ is between 0 and 1. For *Average*, we first show that $|x - \alpha|$ is less than 1, and then show the rest of the integral is 1.

Specifically, all we need to show is that $0 \leq e \leq 1$, where $e = 1 - q = \sqrt{\frac{\int_0^1 x^{r'}(1-x)^{s'}(x-\alpha)^2 dx}{\int_0^1 x^{r'}(1-x)^{s'} dx}}$. Since $(x - \alpha) \leq 1$ when $0 \leq x \leq 1$. Thus we obtain

$\frac{\int_0^1 x^{r'}(1-x)^{s'}(x-\alpha)^2 dx}{\int_0^1 x^{r'}(1-x)^{s'} dx}$

$\leq \frac{\int_0^1 x^{r'}(1-x)^{s'} dx}{\int_0^1 x^{r'}(1-x)^{s'} dx} = 1.$

Since $\frac{\int_0^1 x^{r'}(1-x)^{s'} dx}{\int_0^1 x^{r'}(1-x)^{s'} dx} = 1$, we obtain

$0 \leq e \leq 1.$ □

### 5.2 Monotonicity

We now introduce an important property of trust updates, which we term *monotonicity*. Monotonicity means that for a fixed trust estimate $\overline{\alpha}$, the farther the actual quality of the service provider is from the quality predicted by the estimate the larger is the resulting correction. Here, the correction corresponds inversely to $q$ in Algorithm 1. In other words, monotonicity means that if the error is greater, the correction due to trust update is larger as well.

**Definition 5** *Let a trust update method compute $q_1$ and $q_2$ (corresponding to $\alpha = \alpha_1$ and $\alpha = \alpha_2$, respectively) based on the above description. We define such a method as being* monotonic *if and only if when $\alpha_1 < \alpha_2 < \overline{\alpha}$ or $\overline{\alpha} < \alpha_2 < \alpha_1$, for some trust estimate $\overline{\alpha} \in (0, 1)$, we have*

$$q_1 < q_2$$

Theorem 2 establishes that all of the accuracy measurement definitions that we consider here satisfy monotonicity.

**Theorem 2** *Each of the four definitions of accuracy q as given in Equation 2, Algorithm 4, Equation 3, and Algorithm 5 satisfies monotonicity.*

**Proof**: Linear is trivially seen to be monotonic. To prove that *Max-Certainty* and *Sensitivity* are monotonic, we only need to show that the PCDF function is increasing when $x \in (0, \frac{r}{r+s})$ and decreasing when $x \in (\frac{r}{r+s}, 1)$. We show this by showing that the derivative of the PCDF function is positive when $x \in (0, \frac{r}{r+s})$ and negative when $x \in (\frac{r}{r+s}, 1)$. To prove that *Average* is monotonic, we use Theorem 5 and let $c = \frac{r+1}{r+s+2}$.

More specifically, this theorem is equivalent to showing that $q(\alpha)$ is increasing when $0 < \alpha < \frac{r'}{r'+s'}$ and decreasing when $\frac{r'}{r'+s'} < \alpha < 1$, where $q$ is defined in Equation 3.

By Definition 3, $q(\alpha) = \frac{\alpha^{r'}(1-\alpha)^{s'}}{d}$, where $d = (\frac{r'}{r'+s'})^{r'}(\frac{s'}{r'+s'})^{s'}$. Hence,

$$
\begin{aligned}
q'(\alpha) &= \frac{r'\alpha^{r'-1}(1-\alpha)^{s'} - s'\alpha^{r'}(1-\alpha)^{s'-1}}{d} \\
&= \frac{\alpha^{r'-1}(1-\alpha)^{s'-1}(r'(1-\alpha)-s'\alpha)}{d} \\
&= \frac{\alpha^{r'-1}(1-\alpha)^{s'-1}(r'-\alpha(r'+s'))}{d}.
\end{aligned}
$$

Thus $q'(\alpha) > 0$ when $0 < \alpha < \frac{r'}{r'+s'}$ and $q'(\alpha) < 0$ when $\frac{r'}{r'+s'} < \alpha < 1$.

Hence $q(\alpha)$ is increasing when $0 < \alpha < \frac{r'}{r'+s'}$ and decreasing when $\frac{r'}{r'+s'} < \alpha < 1$.

To prove monotonicity for *Max-Certainty*, it is equivalent to show that $q(\alpha)$ is increasing when $0 < \alpha < \frac{r}{r+s}$ and decreasing when $\frac{r}{r+s} < \alpha < 1$, where $q$ is defined in Algorithm 4. The rest of the proof is as above. □

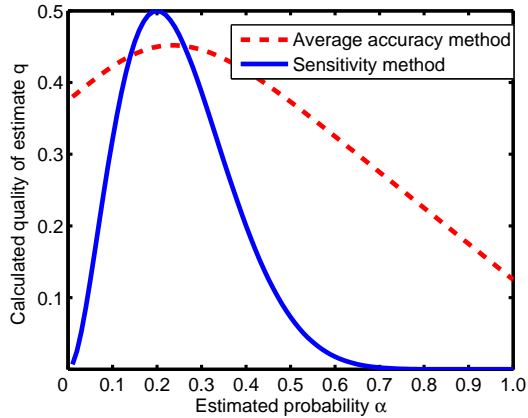

Figure 10: Monotonicity of update methods illustrated: here, each trust update method calculates quality of estimate $q$ given a fixed trust estimate $\langle r', s' \rangle = \langle 2, 8 \rangle$, and $\alpha \in [0, 1.0]$.

Figure 10 shows that *Average* method and *Sensitivity* method satisfy this property. The *Max-Certainty* method is the same as *Sensitivity* method except that it uses $r, s$ instead of $r', s'$. Thus, the figure for *Max-Certainty* would be the same as Figure 10 provided we use $r = 2$ and $s = 8$.

## 5.3 Sensitivity Problems

The property of sensitivity alludes to a problem that some trust update methods face. The idea is that an overly sensitive update method creates unjustifiably large updates. As a result, the trust being placed can oscillate rapidly, leading to near chaotic conditions.

Following Figure 4, Definition 6 specifies what it means for an update method to be asymptotically sensitive. The intuition behind Definition 6 is that as the amount of evidence used to assess the trustworthiness of a source $S$ goes up, it causes potentially erratic updates in the amount of trust placed in the target $R$. Notice that Definition 6 is an *undesirable* property.

**Definition 6** *Let a trust update method compute $q$ based on a referral $\langle r', s' \rangle$ and actual experience $\langle r, s \rangle$. We further assume that $\alpha \neq \alpha'$. We define a trust update method as being asymptotically sensitive if and only if when $\alpha$ and $\alpha'$ are fixed, at least one of the following holds:*

$$\lim_{r'+s' \to \infty} q = 0$$

$$\lim_{r+s \to \infty} q = 0$$

Theorem 3 establishes that *Max-Certainty* and *Sensitivity* satisfy asymptotic sensitivity. This means the above methods are susceptible to sensitivity problems as the amount of evidence to judge a target increases.

**Theorem 3** *Algorithm 4 and Equation 3 satisfy asymptotic sensitivity.*

**Proof sketch**: As for *Max-Certainty* method, let $f(x)$ be the PCDF function. We want to show that $f(\alpha)$ goes to infinity and $f(x)$ goes to zero, where $\alpha = \frac{r}{r+s}$, $x \neq \alpha$. Then $q = \frac{f(\alpha')}{f(\alpha)}$ goes to infinity when $r+s$ goes to infinity and $\alpha' \neq \alpha$.

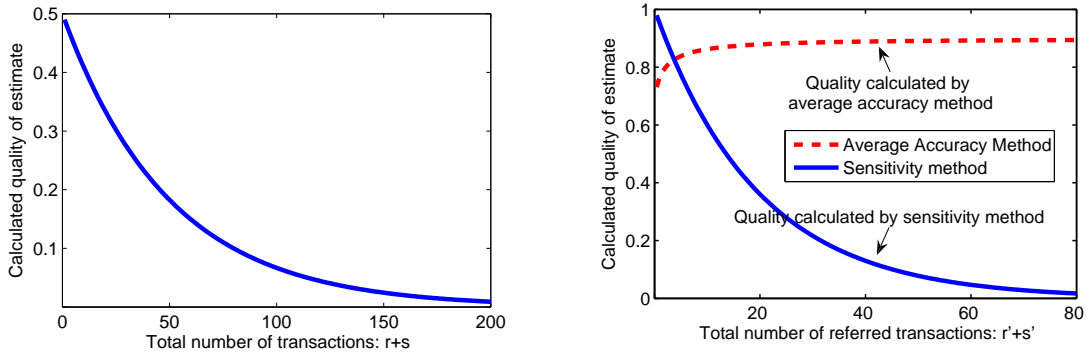

(a) As $r + s$ goes from 1 to 200, the quality calculated by *Max-Certainty* falls.

(b) As $r' + s'$ goes from 1 to 200, the quality calculated by *Sensitivity* falls whereas the quality as calculated by *Average* rises slightly.

Figure 11: Evaluating update methods with respect to quality. In both graphs, $\alpha = 0.6$, $\alpha' = 0.5$.

Figure 11(a) shows that *Max-Certainty* suffers from asymptotic sensitivity as the total number of observations $r+s$ becomes large. And, Figure 11(b) shows that *Sensitivity* suffers asymptotic sensitivity as the total number of transactions in the estimate becomes large.

When $\alpha$ and $\alpha'$ are fixed, both *Average* and *Sensitivity* depend on the total number of transactions $r'+s'$. In Figure 11(b), the quality calculated by *Sensitivity* goes from 1 to 0 and the quality calculated by *Average* goes from about 0.73 to about 0.90. This demonstrates that *Sensitivity* suffers from asymptotic sensitivity since it treats a good estimate as a bad one when $r'+s'$ becomes large, whereas *Average* does not suffer from this problem.

Definition 7 captures the opposite intuition to sensitivity where the accuracy measurement $q$ converges to the difference between the observed and reported probabilities. Theorem 4 shows that *Average*, in contrast with *Max-Certainty* and *Sensitivity*, is not susceptible to sensitivity and is thus more robust than the above methods.

**Definition 7** *Let a trust update method compute $q$ based on the above description. We define such a method as being* convergent *if and only if for a fixed $\alpha$, we have that*

$$\lim_{r'+s'\to\infty} q = 1 - |\alpha - \alpha'|$$

**Theorem 4** *Following Figure 4, when $\alpha$ is fixed, the* Average *method is convergent.*

**Proof sketch**: The convergence of *Average* follows naturally from Theorem 5, given below.

## 5.4 Calculating Average Accuracy

The following formula shows how to calculate *Average*. The important feature of this formula is that it is a closed form for calculating updates. It is exact and does not require computing any integrals, which can be expensive to compute numerically. Hence, in the computational respect too, this method is superior to *Max-Certainty* and *Sensitivity*.

**Theorem 5** *Let $q$ be defined in Algorithm 5. Then*

$$q = 1 - \sqrt{\left(\alpha - \frac{r'+1}{r'+s'+2}\right)^2 + \frac{(r'+1)(s'+1)}{(r'+s'+2)^2(r'+s'+3)}}$$

**Proof sketch**: We need only to show that $\int_0^1 x^{r'}(1-x)^{s'} = \frac{r!s!}{(r+s+1)!}$. We can accomplish this via integration by parts. The boundary terms are zeros. The details are in the Appendix.

To summarize our technical results, we find that all the methods that we consider satisfy the property of the accuracy measure lying within the range $[0, 1]$. *Max-Certainty* and *Sensitivity* satisfy the undesirable property of asymptotic sensitivity, whereas *Linear* and *Average* satisfy the opposite—desirable—property of converging toward the actual measure of accuracy. The advantage of *Average* over *Linear* shows up with respect to its speed of learning, which we demonstrate below through simulation studies.

## 6. Trust Update Scenarios

Now we discuss the two use case scenarios where we apply trust update.

## 6.1 Trust Update for Referrers

Existing trust models lack update methods for an agent to update the extent of trust it places in a referrer, based on the referrals the referrer gives. In general, the trustworthiness of a referrer is best estimated based on how accurate its referrals are.

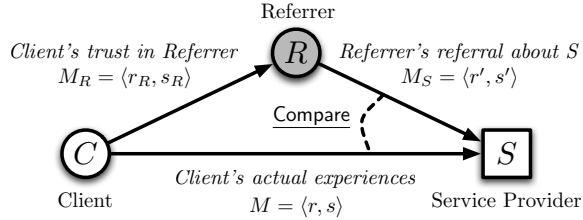

Figure 12: Illustration of trust update for referrers.

The accuracy is determined by comparing the referrals with the observed trustworthiness of the source, as illustrated in Figure 12. This process mirrors exactly the process described for Figure 4, but with the target now being a referrer.

## 6.2 Trust Combination

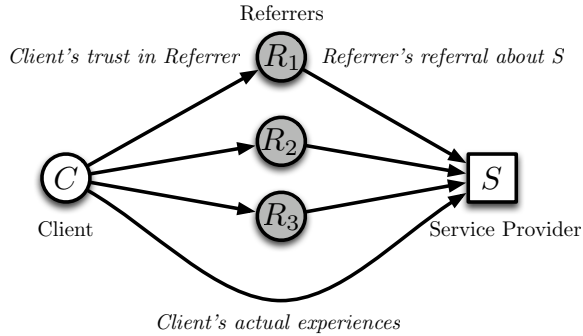

Figure 13: Illustration of trust combination.

After client $C$ has determined the amount of trust it places in the service provider $S$ based on referrals from each of the referrers $R_i$, $C$ consolidates the trust estimates using the propagation operators introduced in Section 2.3. Figure 13 shows a situation with one client, three referrers, and one service provider. $C$ predicts the trustworthiness of the service provider based on the information received from the three referrers. $C$ would make such a prediction when selecting a service provider—typically, before it has obtained sufficient direct experience with $S$.

- $C$ uses the concatenation operator to discount the trust report received from each referrer according to $C$'s trust in that referrer.

- $C$ uses the aggregation operator to combine the discounted trust reports. The combined trust report yields $C$'s estimated trust in the service provider.

### 6.3 Trust Update by Trust in History

To accommodate updating the trust placed in a service provider, we introduce the idea of *trust in history*. We can imagine a "ghost" target reflecting the client's previous level of trust in a specific service provider. The ghost target, in essence, estimates the outcome to be obtained from the service provider. Based on this estimate (and others), the client may estimate its trust in that provider. The client evaluates this ghost target on par with any real referrer.

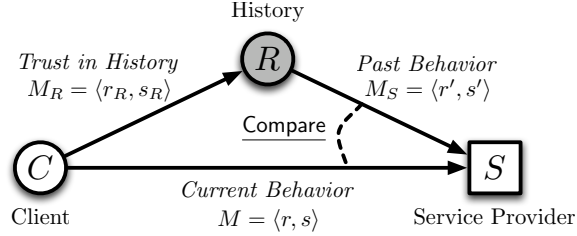

Figure 14: Illustration of trust update by trust in history.

Figure 14 illustrates this scenario as an instantiation of our general model of trust updates, from Section 3. The essential idea is that the client tries to estimate how much trust to place in the past information about a provider. In this manner, we can avoid hard-coding a discount factor to weigh the past information. Instead, the trust placed in history serves as a dynamically computed discount factor that tells us how much to weigh the past.

Algorithm 6 describes a method, *Average-α*, which calculates the discount factor dynamically from trust in history. *Average-α* is similar to *Average-β*. The main difference between the two is that whereas *Average-β* applies to the referrers setting, *Average-α* applies to the trust in history setting (both of these settings are introduced in Section 3).

In Algorithm 6, first, we compare the current behavior $\langle r', s' \rangle$ with the past behavior $\langle r, s \rangle$ to determine how consistent the behavior of the provider $S$ is. If the current behavior $\langle r', s' \rangle$ is close to the past behavior $\langle r, s \rangle$, the trust in history increases; otherwise, it decreases. The closeness is measured by the method **averageAccuracy**, as defined in Algorithm 5. The trust placed in history, $\langle r_R, s_R \rangle$, reflects how static the behavior of $S$ is. Thus, the probability $\alpha$ of trust in history can be used as the discount factor $\beta$, which is high when the new behavior is consistent with the past but low when it is not. Here, we initially set $\langle r_R, s_R \rangle$ to $\langle 0.9, 0.1 \rangle$, i.e., the client trusts its past experience with small amount of confidence. We initially set $\langle r, s \rangle$ to $\langle 0, 0 \rangle$, and we do trust update once the client has done some transactions with the service provider.

A key point of distinction of our approach is that the trust placed in the history $\langle r_R, s_R \rangle$ is not a fixed discount factor; it is based on how much the history matches the subsequent transactions. If the source's behavior changes a lot and cannot be accurately predicted from the history, then the trust placed in the history becomes low, and the historical information is consequently discounted to a greater extent. As a result, the net past evidence that is brought to bear on a prediction goes down in addition to that evidence including more conflict than it would otherwise. Thus the certainty of the resulting prediction is lower than when the new information agrees with the past.

---

**Algorithm 6**: Average-$\alpha$: Yields a discount factor based on $C$'s prior experiences with $S$.

**input** $\langle r, s \rangle$, $\langle r', s' \rangle$, $\langle r_R, s_R \rangle$;

$$\alpha \leftarrow \frac{r}{r + s}$$

$$c \leftarrow \mathbf{c}(r, s)$$

$$c' \leftarrow \mathbf{c}(r', s')$$

$$q \leftarrow \left( 1 - \sqrt{\frac{\int_0^1 x^{r'}(1-x)^{s'}(x-\alpha)^2 dx}{\int_0^1 x^{r'}(1-x)^{s'} dx}} \right)$$

$$r'_R \leftarrow r_R + cc'(1-q)$$

$$s'_R \leftarrow s_R + cc'q$$

$$\beta \leftarrow \frac{r'_R}{r'_R + s'_R}$$

$$r_T \leftarrow r + \beta r'$$

$$s_T \leftarrow s + \beta s'$$

**return** $\langle r_T, s_T \rangle$;

---

## 7. Experimental Evaluation

We now evaluate our approach via simulations to supplement our theoretical analysis. We consider the following main hypotheses in this study.

**Hypothesis 1: Effectiveness** *Average* with trust in history is no worse at prediction than existing approaches for a variety of possible behaviors of service providers (Section 7.2, Section 7.3, Section 7.4, and Section 7.5).

**Hypothesis 2: No tuning** *Average* with trust in history can offer accuracy similar to the traditional approaches without requiring any tuning of parameters (Section 7.4 and Section 7.5).

**Hypothesis 3: Dynamism detection** The certainty computed by *Average* with trust in history reflects the dynamism of service providers. (Section 7.4).

We divide our simulations into two parts. The first part evaluates the effectiveness of our trust update method. Section 7.2 compares our approach with three other models in predicting behavior based on the estimated trustworthiness of referrers. Section 7.3 shows how the trustworthiness estimated by our approach identifies honest from malicious referrers, and further yields accurate reports regarding the service providers.

The second part of our simulation shows the benefits of using trust in history. Section 7.4 compares trust update with and without trust in history in predicting behavior of different profiles. Section 7.5 shows the effectiveness of trust in history on a real dataset from Amazon Marketplace.

We begin in Section 7.1 by introducing some behavior profiles and accuracy metric throughout our evaluation.

## 7.1 Behavior Profiles and Accuracy Metrics

We conduct simulation studies to evaluate our trust update method. To this end, we introduce some interesting behavior profiles for providers to capture a variety of situations that can arise in practice. A *profile* simply means a formal characterization of the behavior of a type of agent. We use the agent profiles to evaluate the effectiveness of the approaches against different kinds of agents.

Table 3: Behavior tracking of different behavior profiles used in Sections 7.2 and 7.4.

| Profile | Example | Behavior Function $X_t$ |
|---|---|---|
| Probability | Amazon ratings | $\begin{cases} 1.0 & 90\% \\ 0.0 & 10\% \end{cases}$ |
| Periodic | Restaurant (lunch and dinner) | $\begin{cases} 1.0 & (\lfloor t/2 \rfloor \bmod 2) \equiv 1 \\ 0.0 & \text{otherwise} \end{cases}$ |
| Damping | Scam artist | $\begin{cases} 1.0 & \text{if } t \leq T/2 \\ 0.0 & \text{otherwise} \end{cases}$ |
| Random | Stock market | $U(0, 1)$ |
| Random Walk | Flight ticket price | $X_{t-1} + \gamma U(-1, 1)$ |
| Momentum | Flight ticket price | $X_{t-1} + \gamma U(-1, 1) + \psi[X_{t-1} - X_{t-2}]$ |

We include the following behavior profiles in our study. Table 3 summarizes these profiles and Figure 15 shows the resulting behaviors schematically. To define the profiles formally, we introduce $X_t$, a *behavior function* to represent the probability of providing a good service at timestep $t$. Here, $U(-1, 1)$ represents the uniform distribution over $[-1, 1]$. The parameters $\psi$ and $\gamma$ are real numbers between 0 and 1. At each timestep $t$, we calculate $X_t$ first, and next determine the quality of service based on $X_t$.

- *Probability* captures providers such as a travel agency or a seller on Amazon Marketplace. For instance, a travel agency might be able to fulfill a passenger's request for a pleasure trip booking with a certain probability.

- *Random* emulates a totally unpredictable service.

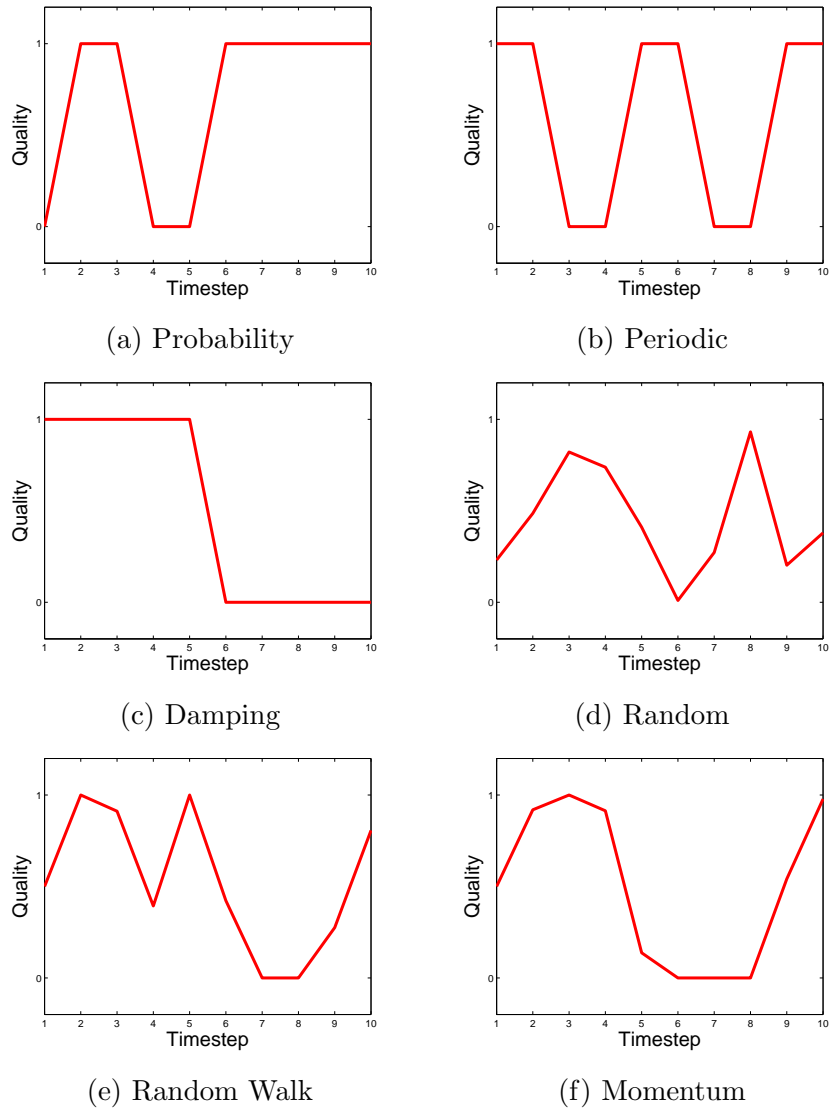

Figure 15: Behavior profiles shown schematically.

- *Periodic* describes a service that changes behavior regularly. For example, a restaurant may employ experienced waiters for dinner, and novices for lunch.

- *Damping* models agents who turn bad after building up their reputation.

- *Random Walk* generalizes over providers whose current behavior depends highly on their immediately previous behavior. For example, the quality of service provided by a hotel would depend upon its recent investments in infrastructure and staff training; thus the next quality of service would show a dependence on the previous quality of service.

- *Momentum* is similar to Random Walk except that its current behavior depends highly on two immediately previous steps.

Among these profiles, Probability, Damping, and Random Walk yield more predictable behaviors than the others because their next outcome relates closely to the previous outcome. Conversely, Random, Periodic, and Momentum are less predictable.

We introduce *average prediction error E* as a measure of the effectiveness of an update method. The idea is that an update method makes a prediction at each timestep and we compare this prediction with the trustworthiness as observed by the client.

**Definition 8** *Let $\langle r'_t, s'_t \rangle$ and $\langle r_t, s_t \rangle$ be the predicted and observed behaviors at timestep $t$. Define $\alpha'_t$ and $\alpha_t$ as usual. Then, the* average prediction error $E$ *over a total of $T$ timesteps equals:*

$$E = \frac{\sum_{t=1}^{T} |\alpha'_t - \alpha_t|}{T}$$

## 7.2 Predicting Referrers of Different Behavior Profiles

We conduct a simulation study to demonstrate the effectiveness of our trust update method. As Table 4 shows, this simulation includes a study of *Average-β* (our approach with fixed discount factor $\beta$) along with three other trust models: *Max-Certainty, Linear-WS*, and *Jøsang*.

Table 4: Trust update methods compared in Section 7.2.

| Update Method | Description | Static Model |
|---|---|---|
| *Linear-WS* (Algorithm 2) | *Linear* with fixed $\beta$ | Wang and Singh |
| *Jøsang* (Algorithm 3) | *Linear* with fixed $\beta$ | Jøsang |
| *Max-Certainty* (Algorithm 4) | *Max-Certainty* with fixed $\beta$ | Wang and Singh |
| *Average-β* (Algorithm 5) | *Average* with fixed $\beta$ | Wang and Singh |

In this experiment, there are 100 timesteps, in each of which the client conducts 50 transactions with the service provider. For concreteness, in our simulations, we set $\langle r_R, s_R \rangle$ and $\langle r, s \rangle$ to $\langle 1, 1 \rangle$ and $\langle 0, 0 \rangle$ in this study. This initial value reflects the intuition that a client might place little trust in a stranger (as a referrer), and has no knowledge of the service provider.

The first simulation compares the *Average-β* trust update method with other methods. In this simulation, there is one client $C$ and one service provider $S$. At each timestep, $C$ obtains a referral from a referrer $R$ about $S$, and itself performs 50 transactions with $S$. Using the various trust update methods, $C$ updates its estimate of the trustworthiness of $R$ based on comparing the referral $R$ gave with $C$'s actual experience. The behavior of the referrer $R$ is defined using each of the above profiles. Note that using different randomness, and $\gamma$ and $\psi$ of Random Walk and Momentum in generating behavior of each profile yields similar results. Here we only show one example of a particular set of profile parameters.

Figure 16 shows the average prediction error of all trust update methods with discount factors of $\beta = 0.00, 0.01, \ldots, 1.00$ for all behavior profiles. For example, for the more predictable profiles such as Probability, Damping, Random Walk, and Momentum, a high

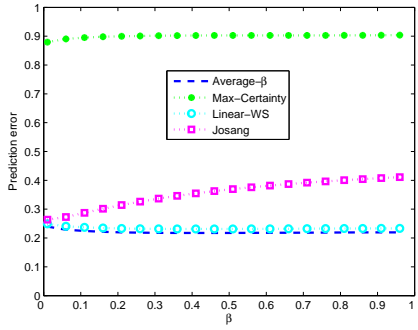

(a) Probability

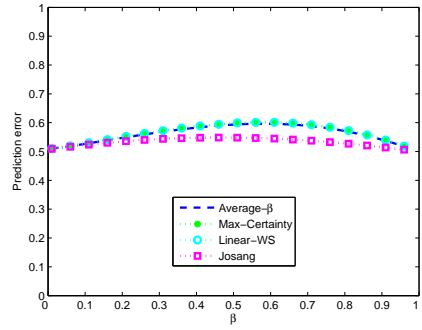

(b) Periodic

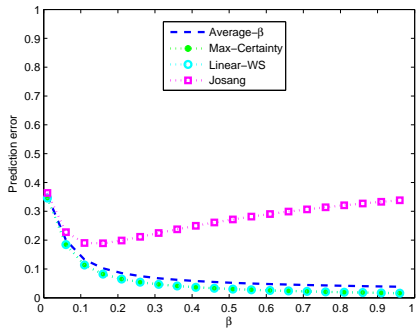

(c) Damping

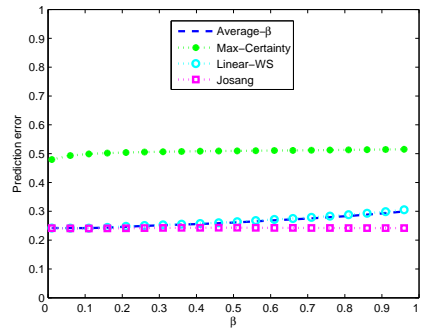

(d) Random

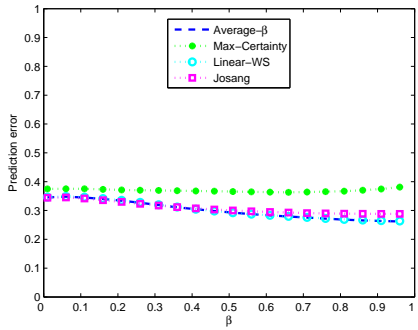

(e) Random Walk

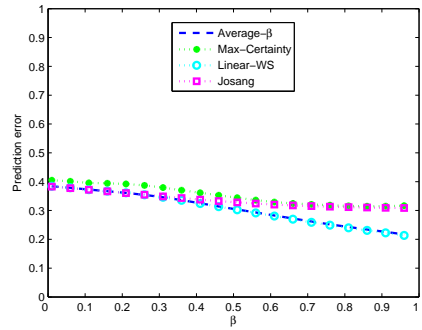

(f) Momentum

Figure 16: Prediction errors with various discount factors. Although *Average-β* does not dominate in all graphs, it yields competitive results, whereas most of the others fail in at least some of the cases.

discount factor $\beta$ yields better predictions because it focuses on recent results, even though it sacrifices a large amount of evidence. Conversely, for the less predictable profiles such as Periodic and Random, it is difficult to determine the best discount factor, because it appears to depend on extraneous factors such as the random seed chosen.

Recall that *Average* is the only one of the update methods that considers certainty. However, this deficiency of *Max-Certainty* and *Linear* is overcome by multiplying certainty in $\langle \delta r_R, \delta s_R \rangle$. As a result, the difference in average prediction errors is not significant.

Section 4.6 describes some cases where *Max-Certainty* fails to predicate accurately. In the context of the Probability and Random profiles, when the probability-certainty density distribution is steep (indicating strong evidence), a small difference between $\alpha$ based on the observed trustworthiness and $\alpha'$ based on the referral can yield a significant punishment in *Max-Certainty*.

To further highlight the advantages of *Average-$\beta$*, we create two new profiles: Rumor and Honest. A Rumor referrer provides accurate reports first but exaggerates the evidence in the second half of the simulation. Suppose the actual experience is $\langle 4, 1 \rangle$. An exaggerated referral might then be, for example, $\langle 40, 10 \rangle$. An Honest referrer provides referrals whose strength depends on its experience. It can accommodate the situation such as at the beginning when it may not have sufficient experience with a provider: an Honest referrer would provide neutral referrals with low certainty.

Figure 17 shows the estimated trustworthiness, respectively, of a referrer following the Rumor and a referrer following the Honest profile. A Rumor referrer provides fair referrals at the beginning and begins to exaggerate later in the simulation. An Honest referrer has provides fair referrals throughout: with little certainty at the beginning and generally with greater certainty later in the simulation. For Rumor, *Average-$\beta$* detects the exaggeration and lowers the trust placed in the referrer accordingly. However, other approaches are not sensitive to this exaggeration. For Honest, *Average-$\beta$* does not punish the referrer even though the referral is inaccurate at the beginning. Once the referrer gathers enough experience and provides good reports, the trust placed in it is built up accordingly. *Max-Certainty* suffers in this case, because, as we discussed in Section 4.6, it punishes Honest for its reports turning out to be false.

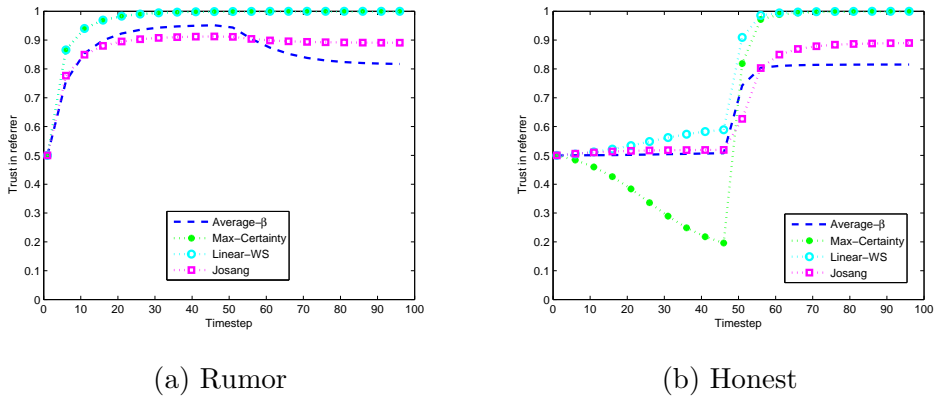

(a) Rumor                                  (b) Honest

Figure 17: Trust (with discount factor $\beta = 0.2$) placed in *Rumor* (exaggerates after timestep 50) and *Honest* (has little knowledge before timestep 50) profiles over time. The result shows our approach *Average-$\beta$* (a) punishes against exaggeration (later in the simulation) and (b) stays neutral about evidence with low confidence (at the beginning of the simulation).

Summary

The foregoing shows that *Average* is effective in evaluating the trustworthiness of referrers of various behavior profiles. *Average* provides competitive predictions against all behavior profiles, whereas other approaches either suffer in some of the profiles or fail to provide accurate predictions. Hence, we conclude that the above supports Hypothesis 1: Effectiveness.

### 7.3 Identifying Robust and Malicious Referrers

Referrers might not be honest or cooperative. This simulation verifies that even if some referrers maliciously provide trust reports indicating a falsely exaggerated amount of evidence, as long as the client can access a few good referrers, it can obtain a good overall estimate of the trustworthiness of the service provider. This experiment involves one client agent, one service provider, and two referrers: one good throughout and one who is good for the first 50 timesteps and then turns bad.

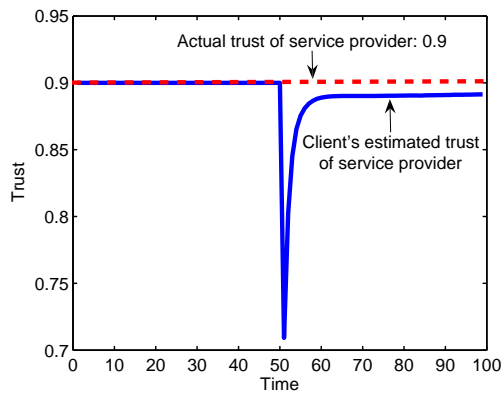

Figure 18: Estimated versus actual trustworthiness of a service provider based on referrals from two referrers, one good throughout and one who is compromised midway.

Figure 18 shows a case where, with just one good referrer to counterbalance one malicious referrer, the client can predict the trustworthiness of the service provider accurately. The service provider offers a good outcome with a probability of 0.90 at all times. The trustworthiness estimated for the service provider is the same as its actual trustworthiness until the $50^{th}$ timestep, when one of the referrers turns bad, as a result of which the estimate drops down to about 0.71. The estimate returns to about 0.88 in about five timesteps and increases slowly back to 0.90.

Figure 19 shows the amount of trust placed in the good and the corrupted referrers. The trust placed in the good referrer begins at about 0.90 and then increases to nearly 1. The trust placed in the corrupted referrer begins from about 0.90 and reaches about 0.98 at the $50^{th}$ timestep, then drops quickly to 0.15 in about ten timesteps. This drop in trust in the corrupted referrer is key to the return to accuracy of the overall assessment of the service provider.

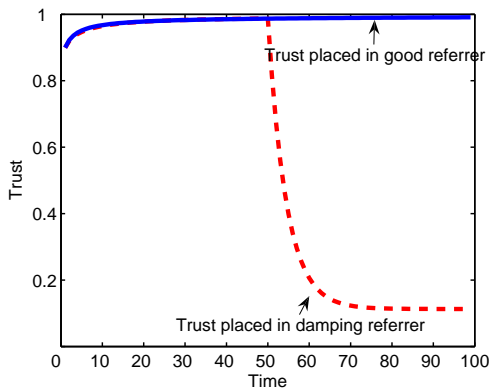

Figure 19: Trust placed in the good referrer versus trust placed in the corrupted referrer.

SUMMARY

The foregoing shows that *Average* is effective in identifying malicious referrers. Besides, *Average* provides accurate predictions despite of erroneous reports from malicious referrers. Hence, we conclude that the above supports Hypothesis 1: Effectiveness.

## 7.4 Predicting Agents of Different Behavior Profiles using Trust in History

Now we evaluate the effectiveness of trust in history in tracking different behaving agents. In this simulation, there is one client $C$ and one service provider $S$. After each 100 timesteps, $C$ performs 50 transactions with $S$. The behavior of the service provider $S$ is defined using the profiles in Table 3. Using three different approaches, we update the estimate of trustworthiness of $S$ and predict its future behavior based on that estimate. In this scenario, we set $\langle r_R, s_R \rangle$ (in Definition 8) to $\langle 0.9, 0.1 \rangle$ in this study. This initial value reflects the intuition that whereas a client would place a fair amount of trust in its own past experience (as history), it might place little trust in a stranger (as a referrer in Section 7.2).

Table 5: Trust update methods compared in Section 7.4.

| Update Method | Description | Static Model |
|---|---|---|
| *Amazon* | No discount | Wang and Singh |
| *Average-$\beta$* (Algorithm 5) | Discount past with fixed $\beta$ | Wang and Singh |
| *Average-$\alpha$* (Algorithm 6) | Discount past with trust in history | Wang and Singh |

Table 5 shows the three approaches compared in this simulation. These approaches discount past experience differently. *Amazon*, like marketplace web sites such as Amazon and eBay, retains all past experience; *Average-$\beta$* decays the past information with a fixed discount factor $\beta$; and, *Average-$\alpha$* decays the past information with a dynamically computed trust in history. For example, suppose $C$ has $\langle 40, 10 \rangle$ and $\langle 25, 25 \rangle$ in first two timesteps. *Amazon* estimates the trustworthiness as $\langle 40 + 25, 10 + 25 \rangle$. *Average-$\beta$* yields $\langle 40\beta + 25, 10\beta + 25 \rangle$, where $\beta$ is a fixed value in $[0, 1]$. We can see *Amazon* is a special case of *Average-$\beta$* when $\beta = 0$. *Average-$\alpha$* uses an adaptive discount factor based on trust in history. Figure 20 shows the average prediction error of all methods with discount factor $\beta = 0.00, 0.01, \ldots, 1$ for all behavior profiles defined in Table 3.

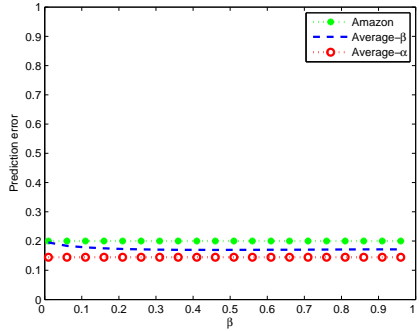

(a) Probability

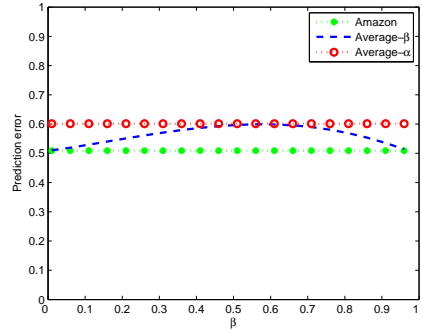

(b) Periodic

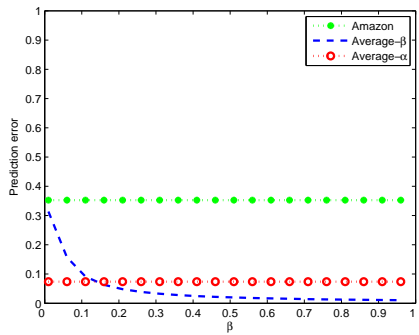

(c) Damping

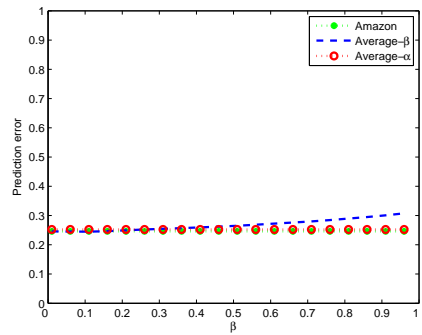

(d) Random

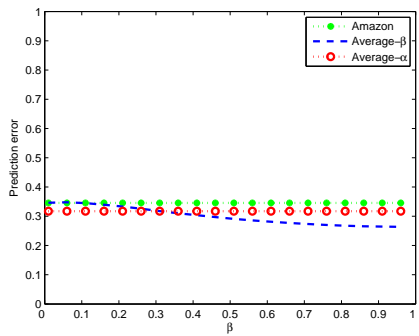

(e) Random Walk

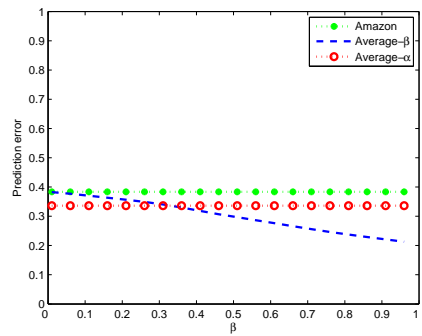

(f) Momentum

Figure 20: Prediction error with various discount factors (lower is better). In each graph, *Amazon* and *Average-$\alpha$* yield horizontal lines since they do not take a discount factor as input.

For Probability, *Average-α* dominates all *Average-β* with different values of *β*. Recall that Probability performs poorly once in a while. When that happens, *Average-α* adapts dynamically by increasing the discount factor. Subsequently, *Average-α* adjusts the discount factor back to a lower value once the behavior becomes more predictable. This adjustment yields better predictions than any fixed discount factor. Note that Probability yields behavior quite similar to real-life agents. An example of these are sellers on Amazon Marketplace, which we discuss in Section 7.5.

As we observed above, some profiles are more predictable than others. For example, for Damping, Random Walk, and Momentum, the next outcome significantly depends upon the previous outcomes. In these cases, discounting old information more by using a high *β* value yields predictions of improved accuracy. However, although *Average-β* with high discount factors provides highly accurate predictions, the high discount factors yield low certainty because the concomitant tendency to consider only the most recent evidence results in reduced evidence, as shown in Figure 21.

Other profiles, especially Random and Periodic, are less predictable. For Random, we need overall information (high *β*) to make the best prediction. For Periodic, higher and lower *β* values yield a lower error than *β* values from the middle, although the error is unacceptably high: the error exceeds 0.50. Note that Periodic changes its quality back and forth every two timesteps. Using *β* = 1 yields perfect prediction when Periodic stays the same but the worst error in the immediately following timestep (because of the alternation of Periodic). Conversely, using *β* = 0 considers the overall behavior: it predicts 0.50 all the time except during the initial several warmup timesteps. All *β* values and *α* yields an error close to 0.50. Periodic is the only case where the behavior cannot be predicted by all approaches. However, Periodic is not realistic in practical cases because a provider who followed it would not gain much utility—*Average-α* can detect its unstable behavior and places low trust in history thus yielding low certainty. The client *C* can react against it accordingly.

Figure 21 compares the certainty of *Amazon*, *Average-α*, and *Average-β* with respect to the *β* that yields the best trust prediction. Recall that certainty reflects two facts: (1) the amount of evidence collected, and (2) the conflict in the evidence. In the first half of the experiment, there is no conflict with the client's own observations. Therefore, for example, in Damping, the certainty of *Average-α* goes up as the history discount decreases (and as the evidence increases). When the referrer turns bad in the middle of the experiment, the certainty drops dramatically, partly because of the conflict in the evidence and partly because, by using a high discount factor to discount the old evidence significantly, *Average-α* in essence reduces the amount of evidence it considers. Conversely, the certainty of *Average-β* is fixed because of the fixed discount factor, except when the conflict occurs in the middle, and its certainty falls down briefly. For profiles such as Random and Periodic, *Average-α* yields lower certainty than *Average-β*, because the behavior is unpredictable. The low certainty of trust prediction can guide the client *C* not to interact with the target because its behavior is unpredictable.

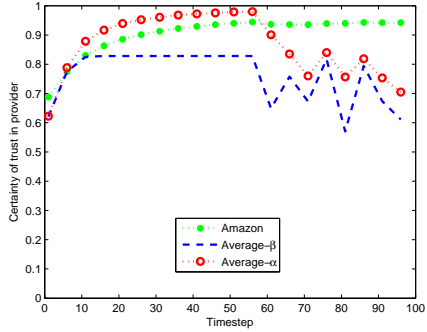

(a) Probability

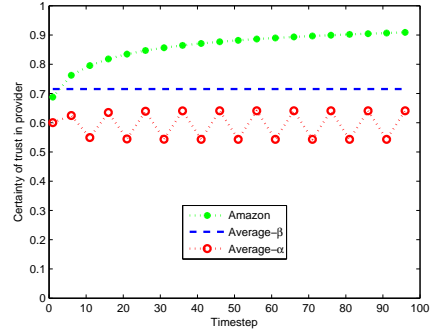

(b) Periodic

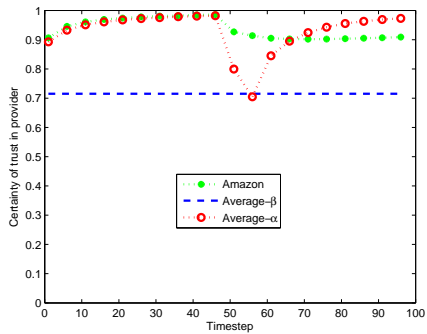

(c) Damping

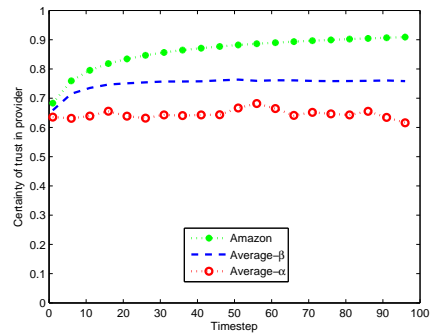

(d) Random

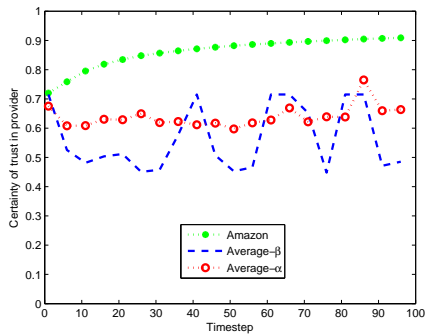

(e) Random Walk

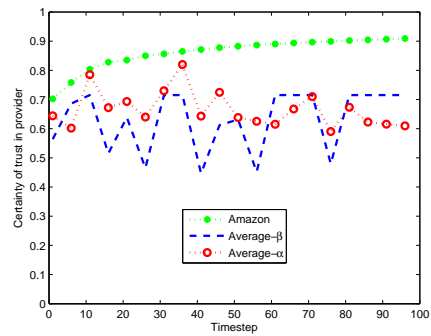

(f) Momentum

Figure 21: Certainty of trust prediction with various discount approaches. Each graph shows the certainty values of *Amazon*, *Average-β* (choosing the β that yields the most accurate prediction), and *Average-α* approaches.

SUMMARY

The foregoing shows that *Average* with trust in history is effective in tracking various dynamic behaviors. *Average* with trust in history provides competitive predictions without prior knowledge of the behaviors and tuning any parameter. Besides, the certainty of *Average* with trust in history can be served as an indicator of the dynamism of the behavior. Hence, we conclude that the above supports Hypothesis 1: Effectiveness; Hypothesis 2: No tuning; and Hypothesis 3: Dynamism detection.

## 7.5 Predicting Amazon Marketplace Data using Trust in History

In order to evaluate the effectiveness of our method in predicting data from the real world, we studied manually collected feedback profiles of five sellers from Amazon. These sellers obtained 60, 107, 180, 235, and 452 feedbacks, respectively. Each feedback is an integer from 1 to 5. We treat each feedback as equaling the quality of service the seller provided during the rated transaction. More precisely, we normalize the rating from $\{1, 2, 3, 4, 5\}$ to $\{0, 0.25, 0.5, 0.75, 1\}$ and treat each normalized rating as the probability obtained from ten transactions. For example, the rating 5 is translated to $\langle 10, 0 \rangle$ and the rating 2 is translated to $\langle 2.5, 7.5 \rangle$. At each timestep, the current feedback can be predicted by using the feedbacks in the past. For example, consider a seller who receives feedbacks 3, 1, 2, and 4, respectively, in the first four timesteps. We can use these feedbacks as a basis for predicting its next feedback. In a simple approach, as supported by Amazon, we can use the average of these feedbacks to predict the fifth feedback. In the above example, we would predict a feedback of $(3 + 1 + 2 + 4)/4 = 2.50$. Suppose at the fifth timestep, the seller actually receives a feedback of 3. Thus our prediction error would be 0.50. The above approach weighs each feedback given in the past equally.

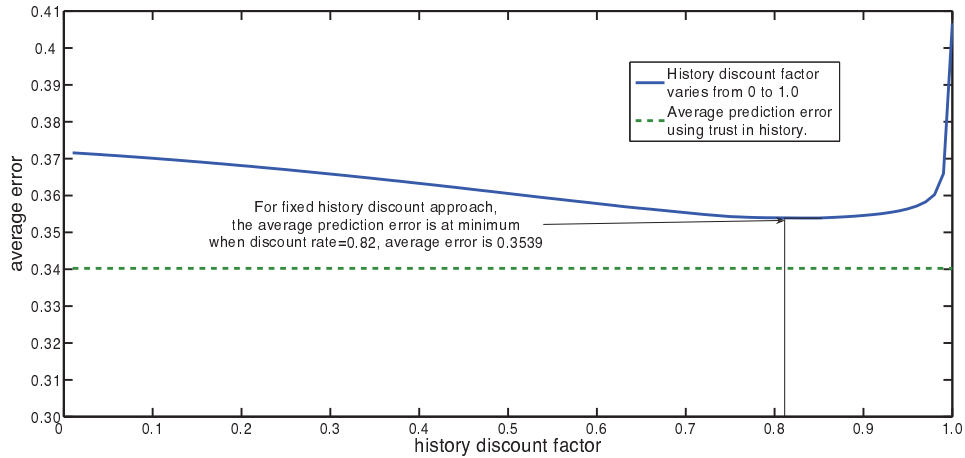

Figure 22: Prediction error of feedback of Amazon seller. Trust in history versus different history discount factors.

Alternatively, we may discount the history by a factor of $\beta \in [0, 1]$. For example, when $\beta = 0.90$, we have $3 \times \beta^3 + 1 \times \beta^2 + 2 \times \beta^1 + 4 \times \beta^0)/(\beta^3 + \beta^2 + \beta + 1)$, which equals $1 \times 0.9^3 + 2 \times 0.9^2 + 3 \times 0.9 + 4)/(0.9^3 + 0.9^2 + 0.9 + 1) = 2.56$. Under this scheme, in the above case, the prediction error would be 0.44. If we use a different discount factor for the history, we would in general obtain a different prediction error. In our experiment, we normalize the feedback to a real number in $[0, 1]$. That is, a feedback of 1, 2, 3, 4, or 5 corresponds to a trust value of 0, 0.25, 0.50, 0.75, or 1, respectively. We compare the prediction error using the trust in history with the prediction error using a specified discount factor. As Figure 22 shows, using the fixed discount in history, when the discount factor is 0.82, the average error of prediction is the lowest, which is 0.35. In general, the error would be higher unless we happened to correctly guess the optimal discount factor. In contrast, using trust in history, the average prediction error is 0.34, which turns out to be lower than using any specific fixed discount factor (*Hypothesis 1*).

An important engineering challenge facing traditional approaches is that, since they require a fixed discount factor for the history as a parameter, we need to manually tune such a discount factor for each application scenario. Such tuning limits the applicability of the traditional approaches substantially. In contrast, our method that uses trust in history automatically adapts to an agent's changing behavior, and thus does not require any manual tuning (*Hypothesis 2*).

Summary

The foregoing shows that *Average* with trust in history is effective in tracking ratings from Amazon without tuning any parameters. Hence, we conclude that the above supports Hypothesis 1: Effectiveness and Hypothesis 2: No tuning.

## 7.6 Summarizing Experimental Results

Our results are twofold. First, in Section 7.2, we show *Average* provides a competitive accuracy measurement on referrals. It can deal with various behavior effectively, detect exaggerated reports, and forgive referrers with no trust information (*Hypothesis 1*). Section 7.3 demonstrates how our trust update method identifies malicious referrer and provides an accurate report from referrals containing false information (*Hypothesis 1*). Second, we show the effectiveness and benefits of trust in history. Section 7.4 presents how trust in history tracks various artificial behavior competitively without parameter tuning (*Hypotheses 1 and 2*). We show that trust in history can preserve a greater amount of evidence than the other approaches and thereby provide additional information about the dynamism of the service provider (*Hypothesis 3*). Section 7.5 shows it works just as well on practical behavior from real datasets (*Hypothesis 1*).

Although *Average-α* is not the most accurate method in all circumstances (*Hypothesis 1*), we consider *Average-α* to be the best solution among these methods. *Average-α* requires no tuning of a discount factor. There are two main advantages for not having to tune $\beta$ by hand.

- Tuning the discount factor can be difficult for a variety of reasons. It is nontrivial to determine a provider's behavior profile. It is difficult to determine the best value of $\beta$ for a specific profile and to maintain that value even as an agent changes its profile dynamically. Using the dynamically changing $\alpha$ as the discount factor can adapt to all kinds of profiles.

- By dynamically tuning the discount factor, *Average-α* can provide dynamic certainty information, which reflects the predictability of the referrers (*Hypothesis 3*). If a provider changes behavior frequently, the certainty computed by the trust in history does not build up. Knowing the certainty may affect an agent's decision-making strategy. Even if the probability is high, the provider may not be trusted, due to a low certainty. Conversely, using β as the discount factor cannot provide such information because it discounts history equally regardless of conflict (*Hypothesis 3*).

## 8. Literature

Trust models have been widely studied (Sabater & Sierra, 2005; Jøsang, Ismail, & Boyd, 2007). Here we focus on some well-known trust models and also the ones that study trust update and evaluate the trustworthiness of referrers based on their referrals.

The Beta Reputation System (BRS) (Jøsang & Ismail, 2002) and SPORAS (Zacharia & Maes, 2000) are two trust models that support the idea of the discount factor. They define a fixed damping factor to control how much past experience should be discounted. This is similar to β, the manually coded discount factor in our experiments. In our approach, the discount factor can be automatically tuned based on how dynamic the agent behavior is. Besides, BRS and SPORAS fail to provide a trust update mechanism to update the estimated trustworthiness of agents based on the accuracy of the trust information they provide.

FIRE (Huynh et al., 2006) and REGRET (Sabater & Sierra, 2002) are two trust models that consider trust information from both individual and social aspects. FIRE estimates trust from four sources: interactions, roles, witnesses, and certified reputations. The trust information in REGRET includes individual and social dimensions. However, both FIRE and REGRET lack a trust update mechanism. Although FIRE can cope with some dynamism, in their experiments, Huynh et al. assume the agent behavior only involves minor changes with an extremely low probability. Our trust update approach copes with various kinds of dynamism. We evaluate our trust update by introducing several dynamic behavior profiles. Our experiments show our trust update provides accurate trust estimation for a variety of natural behavior profiles.

Teacy et al. (2006) develop Travos, one of the trust models based on the beta distribution. Travos calculates trust based on both direct experience and trust information from third parties. Travos also provides a mechanism to measure the accuracy of referrals. Given a referral $\langle r', s' \rangle$ (i.e., a beta distribution), Travos divides the probability density into several disjoint intervals. Suppose the probability α of the actual experience $\langle r, s \rangle$ lies in interval $k$. Then the accuracy of the referral is defined as the probability density ratio of the interval $k$ to all intervals. Similar to *Sensitivity*, their accuracy measurement suffers when the number of transactions is large. Besides, the number of intervals requires human tuning. Teacy et al. suggest that a good trust model should satisfy three requirements: it should provide a comparable trust metric with or without personal experience; it should provide a confidence measure; and it should be able to assess the reliability of trust information sources and discount the information provided by unreliable sources. Our approach satisfies their three requirements of a good trust model. Besides, our approach makes no assumption about agent behavior. However, Travos assumes the agent behavior remains unchanged over time.

Teacy et al. agree that a time-based behavioral strategy is necessary for agents to deal with dynamic behavior. By using our automatically adjusted discount factor, our approach provides a time-based strategy dealing with a variety of dynamic behavior profiles.

Fullam and Barber (2007) study how to choose between trust from direct experience (experience-based) and from referrals (reputation-based). They adopt reinforcement learning to learn a parameter that controls how to aggregate information from experience-based and reputation-based trust. Based on the reward the client gains from the transactions, Fullam and Barber dynamically update the weights of reputation providers (referrers) in a linear manner. Wang and Vassileva (2003) present a Bayesian network-based trust and reputation model for peer-to-peer networks. Their model also treats trust update in a linear manner. They predefine a fixed discount factor to discount past information. Our approach updates trust based on the probability theory. We show how our approach performs better than linear-based trust update approach both theoretically and experimentally.

Ries and Heinemann (2008) propose *CertainTrust*, which is similar to Jøsang's approach. They define trust in terms of the numbers of positive and negative experiences. Their certainty does not reflect the conflict in the evidence, but reflects the amount of evidence. Trust is propagated using two operators: consensus (our aggregation) and discounting (our concatenation). Context dependence is supported by predefining the maximum amount of expected evidence, which is not trivial. Ries and Heinemann update trust in two ways. To update trust from a feedback $f$ (a scalar between $-1$ and 1), they increment the number of positive experiences by $(1+f)/2$ and increment the number of negative experiences by $(1-f)/2$. An alternative is to update the trust placed in an agent based on the accuracy of the recommendations it provides. The accuracy of the recommendations is defined as the *tendency* to the actual behavior. Ries and Heinemann adopt an aging factor analogous to a discount factor. The aging factor normalizes the trust values that exceed the predefined maximum number of experiences. Their aging factor is defined once and used only for normalization. As shown in Section 7, using fixed aging factors poorly deals with various kinds of behavior profiles and these parameters require human tuning. In contrast, our approach can track the dynamic behavior as well as adjust the discount factor based on how dynamic the agents are.

Khosravifar, Gomrokchi, and Bentahar (2009) design a maintenance-based trust model. They define a *timely relevance* factor to discount the past experience when updating trust. The timely relevance factor reflects the time difference between the current time and the time of last update. The amount of discounted information is determined by a domain-dependent variable $\lambda$, which is similar to the discount factor $\beta$ discussed in this paper. When $\lambda$ is high, past experience is forgotten faster. When $\lambda$ is low, trust values tend to consider overall experience. However, Khosravifar et al.'s timely relevance factor requires manual tuning, which is also the common limitation of the trust update methods with the $\beta$ discount factor. The discount factor $\alpha$ in our *Average-*$\alpha$ requires no manual tuning.

Poyraz (Sensoy, Zhang, Yolum, & Cohen, 2009) is a trust-based service selection approach. Poyraz calculates the estimated trustworthiness of web services based on both direct experience and referrals. Poyraz filters out referrals provided by untrustworthy advisors (i.e., referrers). It assesses the trustworthiness of advisors based on private credit and public credit. The private credit is evaluated by comparing the consumer's actual experience with the referral. The comparison produces either a satisfactory or an unsatisfactory assessment based on the consumer's preferences. When the consumer lacks experience, public credit is calculated by comparing the referral with other advisors' referrals. If the referral deviates from the majority, the advisor is considered untrustworthy. Poyraz also provides a time window to discard old trust information. Our approach compares actual experience with referrals based on probability density rather than consumer's preferences. We discount old trust information using a discount factor. Instead of manually adjust the size of the time window, our discount factor is automatically tuned based on how dynamic the agent behavior is.

Paradesi et al. (2009) incorporate trust based on certainty into their work on web service composition. Like our approach, their definition of trust and certainty is based on Wang and Singh's (2007) approach. In their trust framework, Wisp, Paradesi et al. study four types of frequently encountered web service flows in composition. They provide operators to calculate the trust and certainty for the composed web service in each case. Paradesi et al.'s method to update the trust is intuitive based on Wang and Singh and Jøsang's approach: it simply adds up the number of positive and the number of negative transactions separately. Importantly, it does not consider history discount or the effect of aging information, as we have here. In this manner, Paradesi et al.'s work complements our proposed approach, and could be refined to adopt our more sophisticated definitions of trust update.

Mistry, Gürsel, and Sen (2009) estimate reputation scores of sensor nodes based on measurement accuracy in sensor networks. In their framework, a parent node receives reports from its children nodes. Based on the aggregated (average) report, the parent evaluates the trustworthiness of its children. The parent compares the aggregated report and sensed data from each child, and then calculates the error based on the Wilcoxon Signed Rank Test (Wilcoxon, 1945). The reputation of each child reflects the new evidence (the error) based on two update schemes, $\beta$-reputation (Jøsang & Ismail, 2002) and Q-learning (Watkins & Dayan, 1992). Both $\beta$-reputation and Q-learning schemes use a fixed parameter (discount factor $\gamma$ in $\beta$-reputation and learning rate $\alpha$ in Q-learning) to exponentially discount the past evidence, which is similar to our general update (Algorithm 1). Their update schemes require manual tuning and thus lack the ability of dealing with dynamism.

Vogiatzis, MacGillivray, and Chli (2010) build a trust framework on Hidden Markov Models. They apply a probabilistic model to estimate the quality of the service provider with varying behavior. There are some important differences between their approach and ours. Vogiatzis et al. assume that the changes in the behavior of a service provider are slowly varying and can be described as a Wiener process on the quality sequence. A Wiener process is analogous to Brownian motion and supports properties that are not well-motivated when talking about service providers. For example, it requires that the behavior of a service provider have a mean of 0 and a deviation proportional to the time (that is, it should offer the same quality on average as it does at the beginning). Such a model makes sense for a Brownian motion but does not apply to agents who may vary their quality of service due to environmental effects as well as investments in infrastructure, as motivated above. Further, Vogiatzis et al. make additional unjustified assumptions such as that opinions from an honest provider have a normal distribution with a mean of the true quality and that opinions from a dishonest provider have a uniform distribution. Thus, their approach only models agents with randomly decision making, and does not apply to agents who deliberately provide extremely high or extremely low referrals. Overall, the model of Vogiatzis et al. has limited practical applicability in connection with services and agents.

Vogiatzis et al. (2010) evaluate their approach with respect to two types of behaviors: static and damping. We treat dynamism more extensively by defining six dynamic behavior profiles and show how our approach performs against these profiles. However, our definition of reputation of the target (referrer) satisfies the required mathematical properties and it is rigorous. We use heuristics to update the trust of referrals. We have justified our heuristics by proving that our models satisfy important mathematical properties, for example, the farther the referrals are from the believed actual quality of a service provider, the higher is the update ($q$ in Algorithm 1) in the trust placed in the referrer. Our model is computationally efficient. An excessive need for computation is a big shortcoming of Vogiatzis et al.'s and other such traditional approaches. For example, in order to estimate the quality of a service provider with varying behavior, based on 100 transactions, Vogiatzis et al.'s approach would need to calculate multiple integrals with dimension 100. The efficiency would suffer further when calculating the honesty of multiple opinion providers.

Hazard and Singh (2010) identify and axiomatize some common intuitions about trust when viewed from the perspective of the incentives of agents. They relate the trustworthiness of an agent to how it discounts future payoffs: more trustworthy agents have a longer time horizon, an intuition also shared by Smith and desJardins (2009). The above approaches are complementary to ours in that they deal with agents who can strategically alter their behavior whereas we concern ourselves with agents of a fixed type, whose provided quality of service we attempt to estimate. Also, from the incentives perspective, Jurca and Faltings (2007) study mechanisms to ensure agents offer truthful feedback on others. Our approach deals with how to incorporate new evidence in maintaining a trust rating. Their approach applies in settings where one can sanction false reporters and thus promote good behavior.

## 9. Conclusions and Directions

This paper proposes an approach to perform trust updates. It makes the following contributions. One, for the problem of updating the trust placed in referrers on a continuing basis, it develops a mathematically well-justified probabilistic approach for performing updates. Importantly, this approach works on top of a conceptually simple representation for trust that reflects common intuitions about trust and evidence. Further, the proposed approach although cast as a heuristic for calculating trust updates is evaluated (along with some competing heuristics) on mathematical grounds through properties of monotonicity and sensitivity. Two, this paper adapts its referrals approach for updating trust in a provider by modeling trust assessments as referrals from the history of prior interactions. Three, this paper shows that the proposed approach yields performance that compares well with existing approaches *without* requiring any hand tuning of parameters common in previous approaches.

Our investigations have opened up some interesting natural directions for future study. First, an obvious theme is further experimental evaluation. It would be instructive to consider additional types of agents, for instance, discriminative agents. Second, a promising line of inquiry is relating the above to the decision-making strategies for agents to compare trust estimates for service selection. It would be interesting to examine how discounting the past, as we showed above, relates to discounting valuations of the future. Third, in certain settings, especially with widespread sharing of information, updating trust estimates can have significant dynamical effects. Hazard (2010) has studied such dynamical properties of various mechanisms but without explicitly considering referrals. It will be instructive to combine our approach with his.

Fourth, our current definition does not accommodate multivalued events and does not tell us if a referral overestimates or underestimates the quality of the service provider. Multivalued events can be useful in some practical cases. Further, an underestimate might be more desirable than an overestimate—in the former case, you get a pleasant surprise, although it is not always ideal because you would miss out on selecting some good providers because of the underestimation. We have begun to address a suitable representation of trust. However, it would be nontrivial to provide appropriate updating methods.

## Acknowledgments

We thank Chris Hazard and Scott Gerard for helpful comments. Chris provided his dataset for use in our study. This work was partially supported by the U.S. Army Research Office (ARO) under grant W911NF-08-1-0105 managed by the NCSU Secure Open Systems Initiative (SOSI) and partially sponsored by the Army Research Laboratory in its Network Sciences Collaborative Technology Alliance (NS-CTA) under Cooperative Agreement Number W911NF-09-2-0053.

## Appendix A. Proofs of Theorems

**Lemma 6**

$$\int_0^1 x^r(1-x)^s dx = \frac{1}{r+s+1} \prod_{i=1}^r \frac{i}{r+s+1-i}$$

**Proof:** We use integration by parts.

$\int_0^1 x^r(1-x)^s dx = \int_0^1 x^r d(\frac{-1}{s+1}(1-x)^{s+1})$

$= -\frac{x^r(1-x)^{s+1}}{s+1}|_0^1 + \frac{r}{s+1} \int_0^1 x^{r-1}(1-x)^{s+1} dx$

$= \frac{r}{s+1} \int_0^1 x^{r-1}(1-x)^{s+1} dx$

$= \cdots$

$= \frac{r \cdot (r-1) \cdots 1}{(r+s) \cdot (r+s-1) \cdots (s+1)} \int_0^1 (1-x)^{r+s} dx$

$= \frac{1}{r+s+1} \prod_{i=1}^r \frac{i}{r+s+1-i}$. $\qquad\qquad\square$

**Lemma 7** *Given $r$ and $s$ as above, we have that*

$$\lim_{r\to\infty} \sqrt[r]{\prod_{i=1}^r \frac{i}{\alpha r + r + 1 - i}} = \frac{\alpha^\alpha}{(1+\alpha)^{\alpha+1}}$$

*Where $r$ is a positive integer.*

**Proof:** This lemma is used in the next lemma, to show that the right side of an equation approaches a constant, where the equation has duplicated roots, and then the two roots of the equation approach that duplicated root.

$\lim_{r\to\infty} \frac{1}{r} \ln \prod_{i=1}^{r} \frac{i}{\alpha r + r + 1 - i}$

$= \lim_{r\to\infty} \frac{1}{r} \ln(\prod_{i=1}^{r} i \prod_{i=1}^{r} \frac{1}{\alpha r + r + 1 - i})$

$= \lim_{r\to\infty} \frac{1}{r} \ln(\prod_{i=1}^{r} i \prod_{i=1}^{r} \frac{1}{\alpha r + i})$

$= \lim_{r\to\infty} \frac{1}{r} \sum_{i=1}^{i=r} \ln \frac{i}{\alpha r + i}$

$= \lim_{r\to\infty} \frac{1}{r} \sum_{i=1}^{i=r} \ln \frac{\frac{i}{r}}{\alpha + \frac{i}{r}}$

$= \int_0^1 \ln \frac{x}{\alpha + x} dx$

$= \ln \frac{\alpha^\alpha}{(1+\alpha)^{\alpha+1}}$

Therefore,

$\lim_{r\to\infty} \sqrt[r]{\prod_{i=1}^{r} \frac{i}{\alpha r + r + 1 - i}} = \frac{\alpha^\alpha}{(1+\alpha)^{\alpha+1}}.$ □

**Lemma 8** *Let $\alpha = \frac{s}{r}$ and $\alpha$ is fixed. Let $A(r)$ and $B(r)$ be the two values of $x$ that satisfy*

$$\frac{x^r (1-x)^{\alpha r}}{\int_0^1 x^r (1-x)^{\alpha r}} = c \tag{4}$$

*Where $c > 0$.*

$$\lim_{r\to\infty} A(r) = \lim_{r\to\infty} B(r) = \frac{1}{1+\alpha} \tag{5}$$

*Where $r$ is a positive integer.*

**Proof:** The idea is to show that $A(r)$ and $B(r)$ are two roots of an equation $g(x) = \beta(r)$. If $\lim_{r\to\infty} \beta(r) = \beta$ and the equation $g(x) = \beta$ has duplicated roots of $\alpha$, then we have $\lim_{r\to\infty} A(r) = \lim_{r\to\infty} B(r) = \alpha$

$A(r)$ and $B(r)$ are two roots for the equation

$x(1-x)^\alpha = \sqrt[r]{c \int_0^1 x^r (1-x)^{\alpha r} dx}$

since

$\lim_{r\to\infty} \sqrt[r]{c \int_0^1 x^r (1-x)^{\alpha r} dx}$

$= \lim_{r\to\infty} \sqrt[r]{c \frac{1}{\alpha r + r + 1} \prod_{i=1}^{r} \frac{i}{\alpha r + r + 1 - i}}$ (by Lemma 6)

$= \frac{\alpha^\alpha}{(1+\alpha)^{\alpha+1}}$ (by Lemma 7)

$= \frac{1}{1+\alpha}(1 - \frac{1}{1+\alpha})^\alpha$

since $x(1-x)^\alpha$ achieves its maximum at $x = \frac{1}{1+\alpha}$, and $x = \frac{1}{1+\alpha}$ is the only root for the equation

$x(1-x)^\alpha = \frac{1}{1+\alpha}(1 - \frac{1}{1+\alpha})^\alpha$

Therefore,

$\lim_{r\to\infty} A(r) = \lim_{r\to\infty} B(r) = \frac{1}{1+\alpha}.$ □

**Lemma 9** *Let $\alpha = \frac{r}{r+s}$ and $t = r + s$. Let $\alpha$ be fixed and $c \neq \alpha$.*

$$\lim_{t \to \infty} \frac{c^{\alpha t}(1-c)^{(1-\alpha)t}}{\int x^{\alpha t}(1-x)^{(1-\alpha)t}dx} = 0 \tag{6}$$

$$\lim_{t \to \infty} \frac{\alpha^{\alpha t}(1-\alpha)^{(1-\alpha)t}}{\int x^{\alpha t}(1-x)^{(1-\alpha)t}dx} = \infty \tag{7}$$

**Proof:** Let $f(x) = \frac{x^{\alpha t}(1-x)^{(1-\alpha)t}}{\int x^{\alpha t}(1-x)^{(1-\alpha)t}dx}$

Proof of Equation 6:

Without losing generality, assume $0 < c < \alpha$.

For any $\epsilon > 0$, let $A(t)$ and $B(t)$ be defined in Equation 4 where $c = \epsilon$. According to Lemma 8, there is a $T > 0$ such that $c < A(t) < \alpha$ when $t > T$. Since $f(c) < f(A(t)) = \epsilon$, we have $f(c) < \epsilon$. Thus for any $\epsilon > 0$, there is a $T > 0$, such that $\frac{c^{\alpha t}(1-c)^{(1-\alpha)t}}{\int x^{\alpha t}(1-x)^{(1-\alpha)t}dx} < \epsilon$ when $t > T$, which proves Equation 6.

Proof of Equation 7:

For any $N > 0$, let $A(t)$ and $B(t)$ be defined in Equation 4 where $c = 0.50$. Since $f(x) < f(A(t)) = 0.50$ when $x < A(t)$ and $f(x) < f(B(t)) = 0.50$ when $x > B(t)$. Then $\int_0^{A(t)} f(x)dx + \int_{B(t)}^1 f(x)dx < \int_0^1 0.50 dx = 0.50$.

Thus $\int_{A(t)}^{B(t)} f(x)dx = \int_0^1 f(x)dx - \int_0^{A(t)} f(x)dx - \int_{B(t)}^1 f(x)dx$

$1 - \int_0^{A(t)} f(x)dx - \int_{B(t)}^1 f(x)dx > 0.50$ Since $f(x) \leq f(\alpha)$ when $x \in (A(t), B(t))$.

Thus we obtain $(B(t) - A(t))f(\alpha) > 0.50$

According to Lemma 8, there is a $T > 0$ such that $B(t) - A(t) < \frac{1}{2N}$ when $t > T$.

Thus we have $f(\alpha) > 0.50/(B(t) - A(t)) > N$ when $t > T$.

Therefore, for any $N > 0$, there is a $T > 0$ such that $f(\alpha) > N$ when $t > T$, which proves Equation 7. $\square$

**Proof of Theorem 3**

$q = \frac{\alpha^{r'}(1-\alpha)^{s'}}{\alpha'^{r'}(1-\alpha')^{s'}} = \frac{f(\alpha)}{f(\alpha')}$

By Lemma 9, we have $\lim_{t \to \infty} f(\alpha) = 0$ and $\lim_{t \to \infty} f(\alpha') = \infty$, so we have $\lim_{t \to \infty} q = 0$.

The case for *Sensitivity* is the same as the above. $\square$

**Proof of Theorem 4** Let $t = r' + s'$. By Theorem 5, it is equivalent to prove that

$$\lim_{t \to \infty} \sqrt{(\alpha - \frac{r'+1}{r'+s'+2})^2 + \frac{(r'+1)(s'+1)}{(r'+s'+2)^2(r'+s'+3)}} = |\alpha - \alpha'|$$

$\lim_{t \to \infty} \sqrt{(\alpha - \frac{r'+1}{r'+s'+2})^2 + \frac{(r'+1)(s'+1)}{(r'+s'+2)^2(r'+s'+3)}}$

$= \lim_{t \to \infty} \sqrt{(\alpha - \frac{t\alpha'+1}{t+2})^2 + \frac{(t\alpha'+1)(t(1-\alpha')+1)}{(t+2)^2(t+3)}}$

$= |\alpha - \alpha'|$

since $\lim_{t \to \infty} \frac{t\alpha'+1}{t+2} = \alpha'$ and $\lim_{t \to \infty} \frac{(t\alpha'+1)(t(1-\alpha')+1)}{(t+2)^2(t+3)} = 0$ $\square$

**Proof of Theorem 5**

By lemma 6, we have

$\int_0^1 x^{r'}(1-x)^{s'}(x-\alpha)^2 dx$

$= \int_0^1 x^{r'}(1-x)^{s'}(x^2 - 2\alpha x + \alpha^2) dx$

$= \int_0^1 x^{r'+2}(1-x)^{s'} - 2\alpha x^{r'+1}(1-x)^{s'} + \alpha^2 x^{r'}(1-x)^{s'} dx$

$= \frac{(r'+2)!s'!}{(r'+s'+3)!} - 2\alpha \frac{(r'+1)!s'!}{(r'+s'+2)!} + \alpha^2 \frac{r'!s'!}{(r'+s'+1)!}$

So

$q = 1 - \sqrt{\frac{\int_0^1 x^{r'}(1-x)^{s'}(x-\alpha)^2 dx}{\int_0^1 x^{r'}(1-x)^{s'} dx}}$

$= 1 - \sqrt{\frac{\frac{(r'+2)!s'!}{(r'+s'+3)!} - 2\alpha \frac{(r'+1)!s'!}{(r'+s'+2)!} + \alpha^2 \frac{r'!s'!}{(r'+s'+1)!}}{\frac{r'!s'!}{(r'+s'+1)!}}}$

$= 1 - \sqrt{\alpha^2 - 2\alpha \frac{r'+1}{r'+s'+2} + \frac{(r'+1)(r'+2)}{(r'+s'+2)(r'+s'+3)}}$

$= 1 - \sqrt{(\alpha - \frac{r'+1}{r'+s'+2})^2 + \frac{(r'+1)(s'+1)}{(r'+s'+2)^2(r'+s'+3)}}$

$\square$